\documentclass[sn-mathphys-num]{sn-jnl}
\usepackage{graphicx}%
\usepackage{multirow}%
\usepackage{amsmath,amssymb,amsfonts}%
\usepackage{amsthm}%
\usepackage{mathrsfs}%
\usepackage[title]{appendix}%
\usepackage{xcolor}%
\usepackage{textcomp}%
\usepackage{manyfoot}%
\usepackage{booktabs}%
\usepackage{algorithm}%
\usepackage{algorithmicx}%
\usepackage{algpseudocode}%
\usepackage{subcaption}
\usepackage{float}
\usepackage{makecell}

\AddToHook{cmd/appendix/before}{%
	\setcounter{equation}{0}%
	\renewcommand{\theequation}{\thesection.\arabic{equation}}%
	\setcounter{theorem}{0}%
}

\theoremstyle{thmstyleone}%
\theoremstyle{thmstyletwo}%
\theoremstyle{thmstylethree}%

\begin{document}
	
	\title[Article Title]{Lepton Flavor Violation of $Z$ Gauge Boson Decays in Supersymmetric Type-III Seesaw Model}
	
	\author*[1, 2]{\fnm{Vael} \sur{Hajahmad}}\email{whahmad@erzincan.edu.tr}
	\author[3]{\fnm{Murhaf} \sur{Alsayed Ali}} \email{morhaf.alsayed.ali@idlib-university.com}
	\affil*[1]{\orgdiv{Physics Department}, \orgname{Erzincan University}, \country{Turkey}}
	\affil[2]{\orgdiv{Physics Department}, \orgname{Al-Furat University}, \country{Syria}}
	\affil[3]{\orgdiv{Physics Department}, \orgname{Idlib University}, \country{Syria}}

\abstract{In this study, we investigate the lepton flavor violation (LFV) of Z gauge boson decaying into two different flavor charged leptons $Z\rightarrow l_i l_j$ ($Z\rightarrow \tau \mu$, $Z\rightarrow \tau e$ and $Z\rightarrow \mu e$). This work is performed in the framework of the constrained minimal supersymmetric standard model (CMSSM) which is extended by the type-III seesaw mechanism. By considering constraints from the current experimental bounds on neutrino and supersymmetric particle masses, we calculate the branching ratios of the LFV of $Z$ boson decays. The numerical results are found to be $1.30 \times {10}^{-9}$ for both the $\tau \mu$ and $\tau e$ decay channels and $6.40 \times {10}^{-10}$ for the $\mu e$ channel. After applying the constraints from the experimental bounds on the radiative two body decays $l_{i}\rightarrow l_{j} \gamma$, the branching ratios of the LFV of $Z$ boson decays get an additional suppression of $10^{-3}$ for the $\tau \mu$ and $\tau e$ decay channels and $10^{-8}$ for the $\mu e$ channel. Our prediction of the branching ratios is several orders of magnitude below the current experimental bounds.
}




\maketitle
		
\section{Introduction} \label{sec:intro}

The lepton flavor is conserved in the standard model (SM). The observation of neutrino oscillations enables the lepton flavor violation to occur \cite{Esteves_2009}. The SM of elementary particles is a very successful model for explaining of many physical phenomena, which contain a wide range of energy scales that can be covered by the current experiments. The SM cannot be considered as a complete theory because it dose not explain many physical phenomena, like the origin of the neutrino mass which has been discovered by the neutrino oscillation experiments \cite{Primulando2019, PhysRevLett.81.1562, PhysRevLett.89.011301, PhysRevLett.90.021802, PhysRevLett.100.221803}. The mass upper bound in the Karlsruhe tritium neutrino experiment (KATRIN) is estimated to be 1.1 eV ($90\%\ $CL) in 2021 \cite{PhysRevD.104.012005}, later in 2022 it is estimated to be 0.8 eV ($90\%\ $CL) \cite{Aker2022}.

The observation of neutrino oscillations indicates that neutrinos have tiny masses, the SM thus needs to be extended by new mechanisms \cite{Han_2023, Li2020}. The most popular ideas for generating the neutrino mass are seesaw mechanism, which provides a minimal framework to explain the origin of the neutrino mass \cite{Han_2023, Hirsch2012}. The seesaw mechanism can be classified into three primary types: Type-I, type-II and type-III. The type-I and type-III seesaw models introduce three additional singlet (triplet) fermions \cite{MINKOWSKI1977421, yan1979, gell1979supergravity, PhysRevLett.44.912, PhysRevD.22.2227, Foot1989, PhysRevLett.81.1171}, whereas the type-II seesaw model introduces an additional triplet scalar \cite{PhysRevD.22.2227, PhysRevD.25.774, KONETSCHNY1977433, Marshak1980, PhysRevD.22.2860, LAZARIDES1981287, PhysRevD.23.165}.

The study of rare decays like the lepton flavor violation (LFV) is crucial in exploring new physics beyond the standard model (BSM). The LFV decays are of interest because their detection would serve as a clear indication of new physics. The quest to detect such rare decays is ongoing, with investigations being conducted in a variety of processes involving leptons, $Z$ boson, Higgs boson, $Z^\prime$ and various hadrons \cite{Sun_2019, Jurciukonis2022}. 

However, there are many new physical models (BSM) where the branching ratios (BRs) of the LFV decays can be increased to fall within the detectable range of the current experiments. They are grand unified, two Higgs boson doublets, supersymmetric and left-right symmetry models. The predicted BRs in these various extensions of the SM could potentially be detected by the current or future experiments \cite{Sun_2023}. The supersymmetry (SUSY) is a well-motivated solution to the hierarchy problem and the non-baryonic dark matter (DM) problem of the universe. If large hadron collider (LHC) indeed finds signatures of SUSY, then it is extremely appealing to consider the embedding of seesaw mechanism into the supersymmetric framework (SUSY seesaw models). These models perhaps lead to the possibility of having the lepton flavor violation in $Z$ boson, Higgs boson, tau and muon decays \cite{Abada2011}.

Based on the fact that the SM predictions for the $Z$ boson LFV decays ($Z\rightarrow \tau \mu $,  $Z\rightarrow \tau e$ and $Z\rightarrow \mu e$) are suppressed by the neutrino mass. This makes them undetectable by the current experiments at the LHC. By considering the neutrino mass, the SM predictions for the BRs of these decays are estimated to be between ${10}^{-50}$ and ${10}^{-40}$ \cite{altmannshofer2022}. According to the recent results of the ATLAS experiment at the LHC \cite{PhysRevLett.127.271801,PhysRevD.108.032015} and to the expected sensitivity of the future colliders (FCC-ee/CEPC) \cite{Hundi2022, Calibbi2021}, the BRs experimental upper bounds of the $Z$ boson decay channels ($Z\rightarrow \tau e, Z\rightarrow \tau \mu$ and $Z\rightarrow \mu e$) are shown in table \ref{experimental_BR}. 

\begin{table} [h!tbp]
	\centering
	\caption{Experimental upper bounds and expected sensitivity of the BRs of $Z$ boson LFV decays.}
	\label{experimental_BR}
	\begin{tabular}{ccc}
		\hline
		Collider &LHC $(95\% $CL)&FCC-ee/CEPC \\
		\hline
		$BR(Z\rightarrow \tau e)$ &$7.00\times{10}^{-6}$  \cite{PhysRevLett.127.271801} &${10}^{-9}$ \cite{Hundi2022, Calibbi2021} \\	
		$BR(Z\rightarrow \tau \mu)$& $7.20\times{10}^{-6}$ \cite{PhysRevLett.127.271801}&${10}^{-9}$\cite{Hundi2022, Calibbi2021} \\
		$BR(Z\rightarrow \mu e)$&$2.62\times{10}^{-7}$  \cite{PhysRevD.108.032015}&${10}^{-8}-{10}^{-10}$ \cite{Hundi2022, Calibbi2021} \\
		\hline
\end{tabular}
\end{table} 

We focus in this study on the $Z$ boson decays into two different charged lepton flavors $Z\rightarrow l_i l_j$ where $l_i, l_j = e, \mu, \tau$. Our analysis is implemented in the framework of the constrained scenario of the minimal supersymmetric standard model (CMSSM), which is extended by the type-III seesaw model. In the SUSY seesaw model, the large flavor mixings of sleptons induce LFV interactions $l_{i} \bar{l}_{j} V \ (V = \gamma,Z)$. As a result of this, there exists a correlation between the BRs of $Z\rightarrow l_i l_j$ and the radiative two body decays $l_{i}\rightarrow l_{j} \gamma$ \cite {ephjc2004, PhysRevD.67.035004, Dong_2017}. The experimental bounds on both masses of SUSY particles and radiative two body decays $l_{i}\rightarrow l_{j} \gamma$ will be considered to constraint the model parameters, then to evaluate the BRs of $Z\rightarrow l_i l_j$ decays.

Related Studies on $Z$ boson LFV decays in SUSY models have been carried out within the MSSM model with general soft SUSY-breaking term \cite{PhysRevD.67.035004, PhysRevD.40.251} and a minimal CP-violating seesaw model with two heavy Majorana neutrinos \cite{ephjc2004}. The authors in \cite{PhysRevD.40.251} primarily examined how the experimental bounds on the radiative processes $BR(\mu \rightarrow e \gamma \leq 1.7 \times 10^{-10})$ suppress the observation of $(Z\rightarrow \mu e)$ decay. Whereas the authors in \cite{ephjc2004} found that the $Z$ boson LFV decays can still be sizable in supersymmetric seesaw model, among which the largest-rate channel ($Z\rightarrow \tau \mu$) can occur with a branching ratio of $10^{-8}$. In \cite{PhysRevD.67.035004} the $Z\rightarrow l_i l_j$ decays are considered to be uncorrelated from other LFV decays. Thus, for SUSY masses above the bounds the BR$(Z\rightarrow l_il_j) = 2\times{10}^{-9}$ to\ $2\times{10}^{-8}$. After considering the correlation from $l_i\rightarrow l_j \gamma$ decays, at a scenario of independent off-diagonal terms of the three flavors (small mixing angles) the BR$(Z\rightarrow\tau\mu)$ = BR$(Z\rightarrow\tau e)$ = $1.6\times{10}^{-8}$ and the BR$(Z\rightarrow \mu e)$$<1.5\times{10}^{-10}$. Whereas, at large mixing angles scenario the BR$(Z\rightarrow l_il_j)$$\le{10}^{-9}$. 

Additionally, there are another studies, such as minimal 331 model \cite{doi:10.1142/S0217751X11054474}, the B-LMSSM model with R-parity conservation \cite{Dong_2017}, Minimal R-symmetric Supersymmetric Standard Model (MRSSM) \cite{Sun_2019}, the scotogenic model \cite{Hundi2022}, the B-L Supersymmetric Standard Model \cite{Huo_2025} and $U(1)$ extension of the MSSM \cite{PhysRevD.106.055044}. Theses studies have predicted that the BR($Z\rightarrow \tau \mu$) and the BR($Z\rightarrow \tau e$) in the range from ${10}^{-8}$ to ${10}^{-10}$, while the BR($Z\rightarrow \mu e$) is in the range from ${10}^{-13}$ to ${10}^{-19}$.


\section{MSSM-Seesaw Type-III model}

In the type-III seesaw model, the MSSM spectrum of particles is extended by adding a fermionic triplet superfield with zero hypercharge belonging to the adjoint representation of ${SU(2)}_L$ \cite{PhysRevD.83.013003}. It is not sufficient to add one generation of 24-plets to explain all neutrino data if we suppose SU(5) invariant boundary conditions at the grand unified theory (GUT) scale. Therefore, the splitting of the induced mass between different neutrino generations will not be sufficiently large. Thus, we will add three generations of 24-plets. The 24-plets decompose under the standard model gauge group $SU(3)_C \times SU(2)_L \times U(1)_Y$ as follows \cite{Abada2011,PhysRevD.83.013003, PhysRevLett.97.231801,Seesaw3, Hirsch2011}:
\begin{flalign}
{24}_M &= \nonumber {\hat{G}}_M+ {\hat{W}}_M+ {\hat{B}}_M+ {\hat{X}}_M+{\widehat{\bar{X}}}_M \\  
       &=(8,1,0)+(1,3,0)+(1,1,0)+(\bar{3},2,\frac{5}{6})+(3,2,-\frac{5}{6}),
\end{flalign}
the two fermionic components, ${\hat{B}}_M (1, 1, 0)$ and ${\hat{W}}_M (1, 3, 0)$, possess exactly the same quantum numbers as a singlet right-handed neutrino and a fermionic triplet respectively. Thus, if these components are embedded into the SU(5) framework, the type-III seesaw model realization will generally produce a mixture of both type-III and type-I mechanisms \cite{Abada2011,PhysRevD.83.013003,Seesaw3}.
The breaking phase of SU(5) leads the superpotential is below the GUT scale \cite{PhysRevD.83.013003, Seesaw3, BASSO2013698, Hirsch2011, CSABA1996}, the superpotential is defined as follows:
\begin{flalign}
	W = W_{MSSM} + W_{SeesawIII},
\end{flalign}
where $W_{MSSM}$ represents the superpotential for the MSSM model. $W_{SeesawIII}$ represents the superpotential for the SUSY type-III seesaw model. $W_{SeesawIII}$ can be written as follows:  
\begin{flalign}
		\label{eq:Seesaw_III}
W_{seesaw-III}&={\hat{H}}_u\mathrm{(}{\hat{W}}_M\mathrm{\ }Y_W-\ \sqrt{\frac{3}{10}}\mathrm{\ }{\hat{B}}_M Y_B\mathrm{)\ }\hat{L}\mathrm{+\ }{\hat{H}}_u\ \nonumber {\widehat{\bar{X}}}_M Y_X{\hat{d}}_R&\\  
&+\frac{1}{2}{\hat{B}}_M M_B {\hat{B}}_M+\frac{1}{2}{\hat{G}}_M M_G{\hat{G}}_M +\frac{1}{2}{{\hat{W}}_M M}_W{\hat{W}}_M+ {\hat{X}}_M\ M_X{\widehat{\bar{X}}}_M,
\end{flalign}
where $\hat{L}$, $\hat{H}_d$ and $\hat{d}_R$ are superfields of left leptons, up-Higgs boson and right down-quarks. $Y_W, Y_B$ and $Y_X$ represent the Yukawa coupling for fermionic components ${\hat{W}}_M$, $\hat{B}_M$ and $\hat{X}_M$ respectively, while $M_W, M_B$ and $M_X$ represent the mass terms of these fermionic components. At the seesaw scale below the $M_{GUT}$ scale, the Weinberg operator is generated as follows \cite{Abada2011}:
\begin{flalign}
	\label{eq:weinberg operator}
	\frac{1}{2}\kappa_{\nu}\ \hat{L}\ \hat{L}\ {\hat{H}}_u\ {\hat{H}}_u,
\end{flalign}
where $\kappa_{\nu} = \frac{3}{10} Y_{B}^{t} M_{B}^{-1} Y_B + \frac{1}{2} Y_{W}^{t} M_{W}^{-1} Y_W $.\\
The mass matrix of the light neutrino can be written as follows \cite{Abada2011, PhysRevD.83.013003, Seesaw3, Hirsch2011, BASSO2013698}:
\begin{flalign}
	\label{eq:neutrino mass_W_B}
	m_\nu = \frac{-v_u^2}{2}\big[\frac{3}{10} Y_{B}^{t} M_{B}^{-1} Y_B + \frac{1}{2} Y_{W}^{t} M_{W}^{-1} Y_W \big].                  
\end{flalign} 
The type-III formula has two contributions as shown in equation (\ref{eq:neutrino mass_W_B}), one arising from the $SU(2)_L$ singlet ($\propto Y_{B}$) and the other arising from the $SU(2)_L$ triplet ($\propto Y_{W}$) part of the superpotential respectively. Thus, in this situation, the calculation of Yukawa couplings is more complicated. It can be simplified by supposing $Y_B \simeq Y_W$ and $M_B \simeq M_W$ at the seesaw scale. Hence, the equation (\ref{eq:neutrino mass_W_B}) can be written as follows \cite{Abada2011, PhysRevD.83.013003, Seesaw3, Hirsch2011, BASSO2013698}: 
\begin{flalign}
	\label{eq:neutrino mass}
	m_\nu = -\frac{2}{5} v_u^2 Y_{W}^{t} M_{W}^{-1} Y_W.                
\end{flalign}
The equation (\ref{eq:neutrino mass}) depends on a high energy scale ($M_W \sim {10}^{14}$ GeV for $Y_W \sim 1$) \cite{Hirsch2011}. In order to calculate the neutrino mass at low energy scale, we need to know the values of both $M_W$ and $Y_W$ as input parameters at $M_{GUT} \sim {10}^{16}$ GeV. Furthermore, it is obvious from equation (\ref{eq:neutrino mass}) that the $Y_W$ matrix is similar to the neutrino Yukawa matrix $Y_\nu$, because both matrices are diagonalized by the $U^{V}$ lepton mixing matrix. Thus, the structure of $Y_W$ matrix is like the one of $Y_\nu$ matrix \cite{Esteves_2009, Seesaw3, Abada2011, PhysRevD.83.013003}.

We consider the CMSSM model at the GUT scale which is defined by the following parameters:
The universal scalar soft-breaking mass $m_0$, the universal trilinear coupling $A_0$, the universal gaugino soft-breaking mass $m_{1/2}$, the ratio of the Higgs vacuum expectation values tan$\beta$ and the sign of the bilinear $\mu$-term in $W_{MSSM}$, sign($\mu$).\\
We also suppose additional GUT scale SU(5)-motivated boundary conditions for the Yukawa couplings and the mass terms for fermionic components that appearing in equation  (\ref{eq:Seesaw_III}): $Y_W = Y_B = Y_X$ and $M_W = M_B = M_G = M_X$ \cite{Abada2011, Hirsch2011, BASSO2013698}. 

\section{The structure of Yukawa matrix}
The contribution to mixing of the sleptons at one loop is represented by the renormalization group equations (RGEs) \cite{Seesaw3, Abada2011} where the lepton flavor violation occurs in the left-handed sleptons sector, the RGEs can be written as follows:
\begin{equation}
	\label{eq:RGE equations}
	\begin{aligned}
		\Delta m_{\widetilde{L}}^2 & = -\frac{9}{5} \frac{1}{8\ \pi^2}m_{0}^2\left\{3+\frac{A_{0}^2}{m_{0}^2}\ \right\}Y_{W}^\dag Y_W\ log{\left(\frac{M_{GUT}}{M_W}\right)}, \\
		&
		\Delta T_{l}^2 = \frac{-9}{5} \frac{3}{16\ \pi^2}A_0 Y_l Y_{W}^\dag Y_W \ log{\left(\frac{M_{GUT}}{M_W}\right)}, \\
		&
		\Delta m_{\widetilde{l}_R}^2 = 0.
	\end{aligned}
\end{equation}
The presence of Yukawa coupling matrix $Y_W$ in equation (\ref{eq:RGE equations}) induces LFV in the left-handed slepton matrix. In addition, the non-trivial flavor structure of $Y_W$ induces off-diagonal entries in the slepton squared mass matrices. We also notice that the terms of the right-handed sleptons do not receive any contribution to the log-decimal approximation. Furthermore, we notice that the tri-linear coupling terms are suppressed by the charged lepton masses \cite{Abada2011,Seesaw3, PhysRevD.66.075003}.
We need to define the structure of a the Yukawa coupling matrix $Y_{W}$ taking into account the contribution of $Y_{W}$ in the LFV as in equation (\ref{eq:RGE equations}). We also consider the real values of the $Y_W$ matrix to avoid the possible constraints from the moments of the lepton electric dipole ($Y_W^\dag Y_W=Y_W^t\ Y_W $). It is effectively useful and constructive to consider a geometrical interpretation for the Yukawa coupling matrix where its elements are interpreted in the flavor space as components of three generic neutrino vectors ($\mathbf{n}_\mathbf{\mu},\ \mathbf{n}_\mathbf{e}$ and $\mathbf{n}_\mathbf{\tau}$). The Yukawa coupling matrix in the flavor space can be written as follows \cite{Seesaw4, marcano2017}:
\begin{equation}
	\begin{aligned}
		Y_W = \left(\begin{matrix}y_{W11}&y_{W12}&y_{W13}\\y_{W21}&y_{W22}&y_{W23}\\y_{W31}&y_{W32}&y_{W33}\\\end{matrix}\right)\equiv f\left(\mathbf{n}_\mathbf{e}\ \ \ \mathbf{n}_\mathbf{\mu}\ \ \ \mathbf{n}_\mathbf{\tau}\right), 
	\end{aligned}
\end{equation}
where f is the strength of the neutrino Yukawa coupling. $\mathbf{n}_\mathbf{\mu},\ \mathbf{n}_\mathbf{e}$\ and \ $\mathbf{n}_\mathbf{\tau}$ are components of the three generic neutrino vectors in the flavor space.
The term $Y_W^t\ Y_W$ in RGEs equations (\ref{eq:RGE equations}) is related to LFV processes. The term $Y_W^t\ Y_W$ is written as follows:
\begin{equation}
	\label{product_yty}
	\begin{aligned}
		Y_W^t\ Y_W = f\left(\begin{matrix}\mathbf{n}_\mathbf{e}\\\mathbf{n}_
			\mathbf{\mu}\\\mathbf{n}_\mathbf{\tau}\\\end{matrix}\right)f
		\left(\mathbf{n}_\mathbf{e}\ \mathbf{n}_\mathbf{\mu}\ \mathbf{n}_\mathbf{\tau}\right){
			= f}^2\left(\begin{matrix}\left|n_e\right|^2&\mathbf{n}_\mathbf{e}
			\ .\ \mathbf{n}_\mathbf{\mu}&\mathbf{n}_\mathbf{e}\ .\ 
			\mathbf{n}_\mathbf{\tau}\\{\mathbf{n}_\mathbf{\mu}\ .\
				\mathbf{n}}_\mathbf{e}\ &{\mathrm{\ |}n_\mu\mathrm{|} }^2&
			\mathbf{n}_\mathbf{\mu}\ .\ \mathbf{n}_\mathbf{\tau}\\{
				\mathbf{n}_\mathbf{\tau}\ .\ \mathbf{n}}_\mathbf{e}\ &
			\mathbf{n}_\mathbf{\tau}\ .\ \mathbf{n}_\mathbf{\mu}\ &\left|n_
			\tau\right|^2\\\end{matrix}\right),
	\end{aligned}
\end{equation}
where $\mathbf{n}_\mathbf{i}. \mathbf{n}_\mathbf{j} = \left|n_i\right|.\mathrm{|}n_j\mathrm{|}.C_{ij}$. The Parameter
$C_{ij}\equiv$ cos($\theta_{ij}$) is the cosine of three neutrino flavor angles $C_{\tau\mu},\ C_{\mu e}$ and $ C_{\tau e}$. The angle names are induced by the certainty that the cosine of the angle $\theta_{ij}$ controls the transitions of LFV in the $l_i - l_j$ sector. Consequently, the nine input parameters that determine the $Y_W$ matrix can be considered as: Three modulus of the three neutrino vectors ($|n_e|, |n_{\mu}|, |n_{\tau}|$), three relative flavor angles between these vectors ($ \theta_{\mu e}, \theta_{\tau e}, \theta_{\tau \mu}$) and three additional angles ($\theta_1, \theta_2, \theta_3$) which determine the global rotation $\mathcal{O}$ of the three neutrino vectors without changing their relative angles \cite{Seesaw4, marcano2017}. The $Y_W$ matrix values are real, so the $Y_W$ matrix can be written as a product of two matrices: 
\begin{equation}
	Y_W=\ \mathcal{O} \ A,
\end{equation}
where $\mathcal{O}$ is the orthogonal rotation matrix which does not enter in the product $Y_W^{t} Y_W$ ($\mathcal{O}^t \mathcal{O}=I$) therefore it does not affect the study of LFV \cite{Seesaw4, marcano2017}. Furthermore, the elements of the A matrix are determined according to three possible scenarios: tau-electron, tau-muon and muon-electron. For the tau-electron (TE) scenario, we substitute $C_{\mu e}=C_{\tau\mu}=0$ in equation (\ref{product_yty}) where $(\mathbf{n}_\mathbf{\mu},\ \mathbf{n}_\mathbf{e})$ and $(\mathbf{n}_\mathbf{\tau},\ \mathbf{n}_\mathbf{\mu})$ are orthogonal vectors. Thus we get:
\begin{equation}
	\label{eq:yt_y}
	\begin{aligned}
		Y_W^t Y_W={f}^2\left(\begin{matrix}\left|n_e\right|^2&0&\mathbf{n}_\mathbf{e}\ .\ \mathbf{n}_\mathbf{\tau}\\0\ &\left|n_\mu\right|^2&0\\\mathbf{n}_\mathbf{\tau}\ .\ \mathbf{n}_\mathbf{e} &0 &\left|n_\tau\right|^2\\\end{matrix}\right),
	\end{aligned}
\end{equation}
the Yukawa coupling matrix for the electron-tau scenario can be written as follows:
\begin{equation}
	\begin{aligned}
		\label{eq:3}
		Y_{W_{\tau e}} = \mathcal{O}\ A_{\tau e} = \mathcal{O}\ f\left(\begin{matrix}\left|n_e\right|&0&\left|n_\tau\right|C_{\tau e}\\0&\mathrm{\ |}n_\mu\mathrm{|}\ &0\\0&0&\left|n_\tau\right|.\sqrt{1-C_{\tau e}^2}\\\end{matrix}\right). 
	\end{aligned}
\end{equation}
By calculating the product $Y_W^t Y_W$, we get equation (\ref{eq:yt_y}). In this case, the $A_{\tau e}$ matrix is written as follows:
\begin{equation}
	\begin{aligned}
		A_{\tau e} = f \left(\begin{matrix}\left|n_e\right|&0&\left|n_\tau\right|C_{\tau e}\\0&\mathrm{\ |}n_\mu\mathrm{|}\ &0\\0&0&\left|n_\tau\right|.\sqrt{1-C_{\tau e}^2}\\\end{matrix}\right). 
	\end{aligned}
\end{equation}

For the tau-muon (TM) scenario, we substitute $C_{\mu e}=C_{\tau e}=0$ in equation (\ref{product_yty}) where $(\mathbf{n}_\mathbf{\mu},\ \mathbf{n}_\mathbf{e})$ and $(\mathbf{n}_\mathbf{\tau},\ \mathbf{n}_\mathbf{e})$ are orthogonal vectors. Thus, the matrix of Yukawa coupling can be written as follows:
\begin{equation}
	\begin{aligned}
		\label{eq:4}
		Y_{W_{\tau \mu}} = \mathcal{O}\ A_{\tau \mu} = \mathcal{O} f\left(\begin{matrix}\left|n_e\right|&0&0\\0&\mathrm{\ |}n_\mu\mathrm{|}\ &\left|n_\tau\right|C_{\tau\mu}\\0&0&\left|n_\tau\right|.\sqrt{1-C_{\tau\mu}^2}\\\end{matrix}\right). 
	\end{aligned}
\end{equation}
In this case, the $A_{\tau \mu}$ matrix is written as follows:
\begin{equation}
	\begin{aligned}
		A_{\tau \mu} = f\left(\begin{matrix}\left|n_e\right|&0&0\\0&\mathrm{\ |}n_\mu\mathrm{|}\ &\left|n_\tau\right|C_{\tau\mu}\\0&0&\left|n_\tau\right|.\sqrt{1-C_{\tau\mu}^2}\\\end{matrix}\right). 
	\end{aligned}
\end{equation}

While for the muon-electron (ME) scenario, we substitute $C_{\tau \mu}=C_{\tau e}=0$ in equation (\ref{product_yty}) where $(\mathbf{n}_\mathbf{\tau},\ \mathbf{n}_\mathbf{\mu})$ and $(\mathbf{n}_\mathbf{\tau},\ \mathbf{n}_\mathbf{e})$ are orthogonal vectors. Hence, the matrix of Yukawa coupling can be written as follows:
\begin{equation}
	\begin{aligned}
		\label{eq:5}
		Y_{W_{\mu e}} = \mathcal{O}\ A_{\mu e} = \mathcal{O} f\left(\begin{matrix}\left|n_e\right|\ \sqrt{1-C_{\mu e}^2}&0&0\\\left|n_e\right|C_{\mu e}&\mathrm{\ |}n_\mu\mathrm{|}\ &0\\0&0&\left|n_\tau\right|\\\end{matrix}\right).
	\end{aligned}
\end{equation}
In this case, the $A_{\mu e}$ matrix is written as follows:
\begin{equation}
	\begin{aligned}
		A_{\mu e} =  f\left(\begin{matrix}\left|n_e\right|\ \sqrt{1-C_{\mu e}^2}&0&0\\\left|n_e\right|C_{\mu e}&\mathrm{\ |}n_\mu\mathrm{|}\ &0\\0&0&\left|n_\tau\right|\\\end{matrix}\right). 
	\end{aligned}
\end{equation}

The TM scenario ($C_{\tau e} = C_{\mu e} = 0$) may produce large rates for $\tau$-$\mu$ transitions, but always gives negligible contributions to ${\rm LFV}_{\mu e}$ and ${\rm LFV}_{\tau e}$. While the TE scenario ($C_{\tau\mu} = C_{\mu e} = 0$) may gives sizable rates for the $\tau$-e transitions, but always gives negligible contributions to ${\rm LFV}_{\mu e}$ and ${\rm LFV}_{\tau \mu}$. The ME scenario ($C_{\tau\mu} = C_{\tau e} = 0$) may produce large rates only for the $\mu$-e transitions \cite{marcano2017}.
	    \section{The $Z$ boson LFV decays ($Z\rightarrow l_{i} l_{j}$)}
\begin{figure}[h!tbp]
	\centering
	\begin{subfigure}{0.2\textwidth}
		\includegraphics[width=\textwidth]{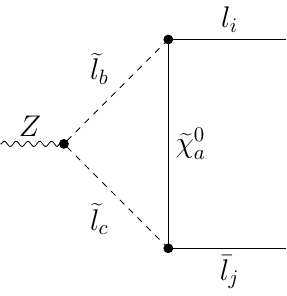}
		\caption{}
	\end{subfigure}
	\hfill 
	\begin{subfigure}{0.2\textwidth}
		\includegraphics[width=\textwidth]{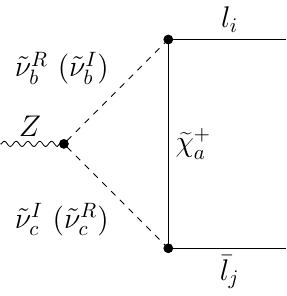}
		\caption{}
	\end{subfigure}
	\hfill
	\begin{subfigure}{0.2\textwidth}
		\includegraphics[width=\textwidth]{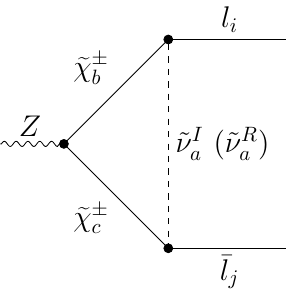}
		\caption{}
	\end{subfigure}	
	\hfill
	\begin{subfigure}{0.2\textwidth}
		\includegraphics[width=\textwidth]{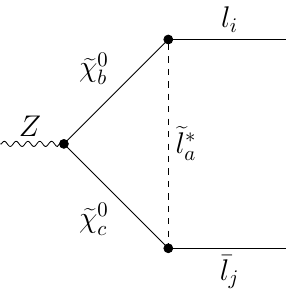}
		\caption{}
	\end{subfigure}
	
	\begin{subfigure}{0.2\textwidth}
		\includegraphics[width=\textwidth]{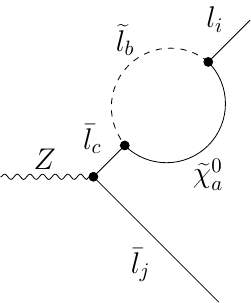}
		\caption{}
	\end{subfigure}
	\hfill
	\begin{subfigure}{0.2\textwidth}
		\includegraphics[width=\textwidth]{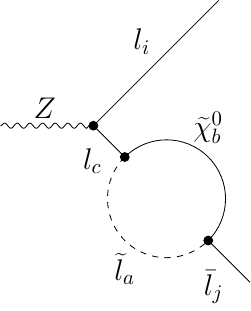}
		\caption{}
	\end{subfigure}
	\hfill
	\begin{subfigure}{0.2\textwidth}
		\includegraphics[width=\textwidth]{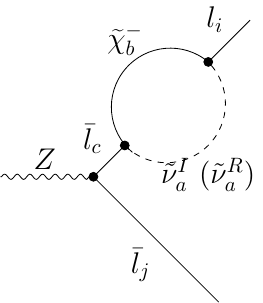}
		\caption{}
	\end{subfigure}
	\hfill
	\begin{subfigure}{0.2\textwidth}
		\includegraphics[width=\textwidth]{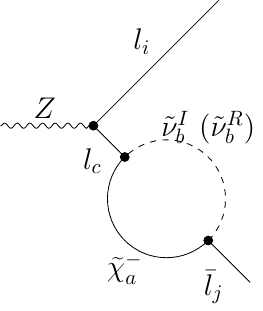}
		\caption{}
	\end{subfigure}
	
	\caption{One loop Feynman diagrams contributing to BR$(Z\rightarrow l_i \bar{l_j})$ in the MSSM-Seesaw type-III model. 
		\label{fig:ZLVF}}
\end{figure}

The flavor mixing of charged sleptons induces the flavor-changing neutral-current couplings $\tilde{\chi}^0 l \tilde{l}$ and $Z \tilde{l}\tilde{l}$, while the flavor mixing of left-handed sneutrinos induces flavor-changing charged-current couplings $\tilde{\chi}^+ l \tilde{\nu}$. These flavor-changing couplings will contribute to the $Z$ boson LFV decays ($Z\rightarrow l_{i}l_{j}$) \cite{Yang2010, ephjc2004}, as shown in Fig.\ref{fig:ZLVF}. The Lagrangian of the $Z$ boson LFV decays can be written as follows \cite {Porod2014, Sun_2019}:
\begin{flalign}
 	\mathcal{L}_{Z l_i l_j} = {\bar{l}}_j [\gamma^\mu (A_1^L P_L+A_1^R P_R)+ p^\mu (A_2^L P_L+A_2^R P_R)] {l}_i Z_\mu,
\end{flalign}
where $P_{L,R}$ = $1/2(1\pm\gamma^5$) are chirality projectors, $l_i$ and $l_j$ represent the lepton flavors, $p$ is the 4-momentum for $l_j$ and $Z_\mu$ is the Higgs boson field. The coefficients $A_1^L, A_1^R, A_2^L, A_2^R$ can be obtained from the amplitudes of Feynman diagrams as shown in Fig.\ref{fig:ZLVF}.
By neglecting the masses of charged leptons we can write the branching ratio equation of the $Z$ boson LFV decays as follows \cite{PhysRevD.63.096008, Porod2014, Sun_2019}:
\begin{flalign}
	BR (Z\rightarrow l_{i} l_{j})=
	BR(Z\rightarrow l_i \bar{l_j})+BR(Z\rightarrow \bar{l_i} l_j) = \frac{{\Gamma(Z\rightarrow l_i \bar{l_j)}}+{\Gamma(Z\rightarrow \bar{l_i} l_j)}}{\Gamma_Z},
\end{flalign}
where $\Gamma_Z$ represents the total decay width of $Z$ boson ($\Gamma_Z = 2.4952$ GeV) \cite{PhysRevD.110.030001}, while the decay width is given by \cite{PhysRevD.63.096008, Porod2014}:
\begin{flalign}	
	\Gamma(Z\rightarrow l_{i} l_{j}) = \frac{m_Z\ }{48\pi} \big[2(\left|A_1^L\right|^2 +\left|A_1^R\right|^2)+\frac{m_Z^2}{4}(\left|A_2^L\right|^2+\left|A_2^R\right|^2) \big].	
\end{flalign}
Thus the final relation of the branching ratio can be written as follows:
\begin{flalign}	
	BR (Z\rightarrow l_{i} l_{j})& = \frac{m_Z\ }{48\pi \Gamma_Z} \big[2(\left|A_1^L\right|^2 +\left|A_1^R\right|^2)+\frac{m_Z^2}{4}(\left|A_2^L\right|^2+\left|A_2^R\right|^2) \big],	
\end{flalign}
where the coefficients $A_1^{L/R}$ and $A_2^{L/R}$ are combinations of the corresponding coefficients to each Feynman diagram as in Fig.\ref{fig:ZLVF}. They can be expressed as follows:
\begin{flalign}
A_1^{L/R} = A_{1a}^{L/R}+A_{1b}^{L/R}+A_{1c}^{L/R}+A_{1d}^{L/R}+A_{1e}^{L/R}+A_{1f}^{L/R}+A_{1g}^{L/R}+A_{1h}^{L/R}
\end{flalign}
\begin{flalign}
A_2^{L/R} = A_{2a}^{L/R}+A_{2b}^{L/R}+A_{2c}^{L/R}+A_{2d}^{L/R}+A_{2e}^{L/R}+A_{2f}^{L/R}+A_{2g}^{L/R}+A_{2h}^{L/R}
\end{flalign}
The contributions of neutralino-slepton loops are derived form Fig.\textcolor{blue}{\ref{fig:ZLVF}(a, d, e, f)}, while the contributions of chargino-sneutrino loops are derived from Fig.\textcolor{blue}{\ref{fig:ZLVF}(b, c, g, h)}. The diagrams with sneutrinos in Fig.\textcolor{blue}{\ref{fig:ZLVF}(b)} or neutralinos in Fig.\textcolor{blue}{\ref{fig:ZLVF}(d)} are do not couple to the photon and do not contribute to $l_{i}\rightarrow l_{j} \gamma$ decays. The diagrams (e, f, g, h) in Fig.{\ref{fig:ZLVF}} are not relevant to the photon processes since they do not give dipole contributions \cite{PhysRevD.67.035004}.
The Feynman diagrams contributions from Fig.\textcolor{blue}{\ref{fig:ZLVF}(a, b)} are expressed as follows:
\begin{align}
	\label{40} 
	A_1^{L(a, b)} =  2 V_1^L\ V_2^R \ V_Z C_{00},  
\end{align}
\begin{align} 
	\label{41}
	A_2^{L(a, b)} = 2 V_1^L V_2^L V_Z (C_0 + C_1 + C_2) M, 
\end{align} 
\begin{align}
	\label{42} 
	A_1^R = A_1^L(L\leftrightarrow R), \ A_2^R = A_2^L(L\leftrightarrow R).
\end{align} 
The couplings corresponding to Fig.\textcolor{blue}{\ref{fig:ZLVF}(a)} are: $V_1^L = \Gamma^{\tilde{\chi}^0 l \tilde{l}^*,L}_{a, i, b}$, $V_2^{L/R} = \Gamma^{\bar{l}\tilde{\chi}^0 \tilde{l} ,\ L/R}_{j, a, c}$ and $V_Z = \Gamma^{\tilde{l}\ \tilde{l}^*Z}_{c, b}$. From equation (\ref{41}), the M parameter represents the neutralino mass ($M=m_{\tilde{\chi}^0_{{a}}}$). Table \ref{coupling_vertex} shows the definition of the coupling constants for the Feynman diagrams.

The couplings corresponding to Fig.\textcolor{blue}{\ref{fig:ZLVF}(b)} are as following: $V_1^L = \Gamma^{\tilde{\chi}^+ l \tilde{\nu}^{I/R},L}_{a, i, b}$, $V_2^{L/R} = \Gamma^{l \tilde{\chi}^- \tilde{\nu}^{I/R} ,\ L/R}_{j, a, c} $ and $V_Z = \Gamma^{\tilde{\nu}^I \tilde{\nu}^R Z }_{b, c} = -\Gamma^{\tilde{\nu}^I \tilde{\nu}^R Z }_{c, b}$. Regarding Fig.\textcolor{blue}{\ref{fig:ZLVF}(a, b)}, the parameters $C_0, C_{00}, C_1$ and $C_2$ denote the standard three-point functions \cite{PASSARINO1979151, HAHN2000231}, they can be calculated using a Mathematica package called the Package-X \cite{PATEL2015276}. The arguments of C functions for Fig.\textcolor{blue}{\ref{fig:ZLVF}(a, b)} are ($0, m^2_{Z}, 0, m^2_{\tilde{\chi}^0_{{a}}}, m^2_{\tilde{l}_{{c}}}, m_{\tilde{l}_{{b}}}$) and ($0, m^2_{Z}, 0, m^2_{\tilde{\chi}^-_{{a}}}, m^2_{\tilde{\nu}^{R/I}_{{c}}}, m^2_{\tilde{\nu}^{I/R}_{{b}}}$) respectively, with the external fermion masses have been set to zero. 
The contributions obtained from Fig.\textcolor{blue}{\ref{fig:ZLVF}(c, d)} are expressed as follows:
\begin{align} 
	\label{43}
	A_1^{L(c, d)} = V_1^L\ V_2^R \ \Big[V_Z^{L} C_{0} m_1 m_2 -V_Z^{R}(B_0-2C_{00}+C_0) m_3^{2}\Big],
\end{align}
\begin{align} 	
	\label{44}
	A_2^{L(c, d)} = 2 V_1^L\ V_2^L \ \Big[-V_Z^{L} C_{1} m_1 +V_Z^{R}(C_0+C_{1}+C_2) m_2^{2}\Big],
\end{align}
\begin{align} 
	\label{45}
	A_1^R =	A_1^L(L\leftrightarrow R), \  A_2^R = A_2^L(L\leftrightarrow R).
\end{align}
\begin{table} [h!tbp]
	\centering
	\caption{The definition of the coupling constants for the Feynman diagrams. The concrete forms of these couplings are available in the Appendix \ref{section:Appendix-A Vertexes}.}
	\label{coupling_vertex}
	\begin{tabular}{ll}
		\hline
		Coupling Constant & Coupling Definition \\
		\hline \\ 
		$\Gamma^{\tilde{\chi}^0 l \tilde{l}^*,L}_{a, i, b}$ & Coupling of neutralino-lepton-slepton vertex \\ 
		\\
		$\Gamma^{\bar{l}\tilde{\chi}^0 \tilde{l} ,\ L/R}_{j, a, c}$ & Coupling of anti lepton-neutralino-slepton vertex \\
		\\
		$ \Gamma^{\tilde{l}\ \tilde{l}^*Z}_{c, b}$ & Coupling of slepton-slepton-$Z$ boson vertex \\
		\\
		$\Gamma^{\tilde{\chi}^+ l \tilde{\nu}^{I/R},L}_{a, i, b}$& Coupling of chargino-lepton-(CP-odd/even) sneutrino vertex \\
		\\
		$\Gamma^{l \tilde{\chi}^- \tilde{\nu}^{I/R} ,\ L/R}_{j, a, c} $& Coupling of the lepton-chargino-(CP-odd/even) sneutrino vertex \\
		\\
		$\Gamma^{\tilde{\nu}^I \tilde{\nu}^R Z }_{c, b}$ &Coupling of CP-odd(even) sneutrino - $Z$ boson vertex \\
		\\
		$\Gamma^{\tilde{\chi}^+ \tilde{\chi}^- Z,{L/R}}_{c, b}$& Coupling of chargino-chargino-$Z$ boson vertex \\
		\\
		$\Gamma^{\tilde{\chi}^0 \tilde{\chi}^0 Z,{L/R}}_{c, b}$& Coupling of neutralino-neutralino-$Z$ boson vertex \\
		\\
		$\Gamma^{\bar{l}l Z ,L}$ & Coupling of anti lepton-lepton-$Z$ boson vertex \\
		\hline \\
	\end{tabular}
\end{table}
The couplings corresponding to Fig.\textcolor{blue}{\ref{fig:ZLVF}(c)} are defined as follows: $V_1^{L/R} = \Gamma^{\tilde{\chi}^+ l \tilde{\nu}^{I/R},{L/R}}_{b, i, a}$, $V_2^{R/L} = \Gamma^{\bar{l} \tilde{\chi}^- \tilde{\nu}^{I/R},{R/L}}_{j, c, a}$ and $V_Z^{L/R} = \Gamma^{\tilde{\chi}^+ \tilde{\chi}^- Z,{L/R}}_{c, b}$. For Fig.\textcolor{blue}{\ref{fig:ZLVF}(d)}, the couplings are as follows: $V_1^{L} = \Gamma^{\tilde{\chi}^0 l \tilde{l}^*,{L}}_{b, i, a}$, $V_2^{R/L} = \Gamma^{\bar{l} \tilde{\chi}^0 \tilde{l},{R/L}}_{j, c, a}$ and $V_Z^{L/R} = \Gamma^{\tilde{\chi}^0 \tilde{\chi}^0 Z,{L/R}}_{c, b}$. 

For Fig.\textcolor{blue}{\ref{fig:ZLVF}(c)}, the parameters $m_1, m_2$ and $m_3$ represent the masses of chargino and (CP-odd/CP-even) sneutrino ($m_1=m_{\tilde{\chi}^-_{{b}}}, m_2=m_{\tilde{\chi}^-_{c}}, m_3=m_{\tilde{\nu}^{I/R}_{a}}$). Whereas for Fig.\textcolor{blue}{\ref{fig:ZLVF}(d)}, $m_1$ and $m_2$ represent the mass of neutralino ($m_1=m_{\tilde{\chi}^0_{{b}}}, m_2=m_{\tilde{\chi}^0_{c}})$ and $m_3$ represents the mass of slepton $(m_3=m_{\tilde{l}_{a}}$). The arguments of C functions for Fig.\textcolor{blue}{\ref{fig:ZLVF}(c, d)} are ($m^2_{Z},0 , 0, m^2_{\tilde{\chi}^-_{{c}}},m^2_{\tilde{\chi}^-_{{b}}}, m^2_{\tilde{\nu}^{I/R}_{{a}}}$) and ($m^2_{Z},0 , 0, m^2_{\tilde{\chi}^0_{{c}}}, m^2_{\tilde{\chi}^0_{b}}, m^2_{\tilde{l}_{a}}$) respectively, with the external fermion masses have been set to zero. The $B_0$ denotes a two-point function. The arguments of $B_0$ function from  Fig.\textcolor{blue}{\ref{fig:ZLVF}(c, d)} are ($m^2_{Z}, m^2_{\tilde{\chi}^-_{{b}}}, m^2_{\tilde{\chi}^-_{{c}}}$) and ($m^2_{Z}, m^2_{\tilde{\chi}^0_{{b}}}, m^2_{\tilde{\chi}^0_{{c}}}$) respectively.

The contributions of Feynman diagrams in Fig.\textcolor{blue}{\ref{fig:ZLVF}(e, f, g, h)} (for which the $Z$ boson does not couple to a scalar particle, hence $A_2^{L/R}$ = 0) can be written as follows:
\begin{align} 
	\label{46}
	A_1^{L} = \frac{V_Z^L}{m_1^2-m_2^2}\big[-V_1^L V_2^R B_1 m_1^2+V_1^R V_2^R B_0 m_1 m_3-V_1^R V_2^L B_1 m_1 m_2+V_1^L V_2^L B_0 m_3 m_2\big],
\end{align}
\begin{align} 
	\label{47}
	A_1^{R} \ = & \ \  A_1^{L}(L\leftrightarrow R).
\end{align}
The couplings corresponding to Fig.\textcolor{blue}{\ref{fig:ZLVF}(e, f)} are defined as follows:  $V_1^{L/R} = \Gamma^{\tilde{\chi}^0 l \tilde{l}^*,L/R}_{a, i, b}=\Gamma^{\tilde{\chi}^0 l \tilde{l}^*,L/R}_{b, c, a}$, $V_2^{R/L} = \Gamma^{\bar{l} \tilde{\chi}^0 \tilde{l},R/L}_{a, i, b}=\Gamma^{\bar{l} \tilde{\chi}^0 \tilde{l},R/L}_{j, b, a}$ and $V_Z^L = \Gamma^{\bar{l}l Z ,L}_{j, c}=\Gamma^{\bar{l}l Z ,L}_{c, i}$. For Fig.\textcolor{blue}{\ref{fig:ZLVF}(e)}, the parameters $m_1, m_2$ and $m_3$ represent the masses of both lepton and neutralino ($m_1=m_{{l}_{i}}, m_2=m_{{l}_{c}}, m_3=m_{\tilde{\chi}^0_{{a}}}$). Similarly, for Fig.\textcolor{blue}{\ref{fig:ZLVF}(f)}, the masses are defined as $m_1=m_{{l}_{j}}, m_2=m_{{l}_{c}}$ and $m_3=m_{\tilde{\chi}^0_{{b}}}$.

The couplings corresponding to Fig.\textcolor{blue}{\ref{fig:ZLVF}(g, h)} are given by: $V_1^{L/R} = \Gamma^{\tilde{\chi}^+ l \tilde{\nu}^{I/R},L/R}_{b, i, a}=\Gamma^{\tilde{\chi}^+ l \tilde{\nu}^{I/R},L/R}_{a, c, b}$, $V_2^{R/L} = \Gamma^{\bar{l} \tilde{\chi}^- \tilde{\nu}^{I/R},R/L}_{c, b, a}=\Gamma^{\bar{l} \tilde{\chi}^- \tilde{\nu}^{I/R},R/L}_{j, a, b}$ and $V_Z^L = \Gamma^{\bar{l}l Z ,L}_{j, c}=\Gamma^{\bar{l}l Z ,L}_{c, i}$. In Fig.\textcolor{blue}{\ref{fig:ZLVF}(g, h)}, the parameters $m_1, m_2, m_3$ are the masses of leptons and chargino. They are defined as ($m_1=m_{{l}_{i}}, m_2=m_{{l}_{c}},  m_3=m_{\tilde{\chi}^- _{{b}}})$ and ($m_1=m_{{l}_{j}}, m_2=m_{{l}_{c}},  m_3=m_{\tilde{\chi}^- _{{a}}}$) respectively.The arguments of B functions for Fig.\textcolor{blue}{\ref{fig:ZLVF}(e, f, g, h)} are ($m^2_{{l}_{i}}, m^2_{\tilde{\chi}^0_{{a}}}, m^2_{\tilde{l}_{{b}}}$), ($m^2_{{l}_{j}}, m^2_{\tilde{\chi}^0_{{b}}}, m^2_{\tilde{l}_{{a}}}$), ($m^2_{{l}_{i}}, m^2_{\tilde{\chi}^- _{{b}}}, m^2_{\tilde{\nu}^{I/R}_{{a}}}$) and ($m^2_{{l}_{j}}, m^2_{\tilde{\chi}^- _{{a}}}, m^2_{\tilde{\nu}^{I/R}_{{b}}}$) respectively.

		\section{Numerical Results and Discussion}
In this section, the numerical results are implemented using the SARAH, SPheno and FlavorKit packages at full two-loops RGEs \cite{Sarah1, Sarah2, Sarah3, Bernigaud2022}. According to the considered CMSSM-Seesaw type-III model, the final parameters in this study are as the following:
$Y_W$, $\ M_W,\ f,\ \left|n_e\right|,\ \mathrm{|}n_\mu\mathrm{|},\ \left|n_\tau\right|,\ C_{\tau\mu},\ C_{\tau e},\ C_{\mu e},\ m_{1/2},\ m_0$,\ $A_0$, tan$\beta$ and sign($\mu$).\\
In our calculations the soft symmetry breaking terms are constrained by several theoretical and experimental conditions. Such as the conservation of R-parity and the lightest SUSY particle of the studied model is the neutralino. Furthermore, the SUSY particle masses (charginos, neutralinos, sleptons and sneutrinos) which are calculated with SPheno must be above the recent experimental lower mass bounds \cite{PhysRevD.110.030001}, as shown in table \ref{tab:Sparticles limits}. 
\begin{table}[h!tbp]
	\centering
	\caption{Experimental mass bounds on SUSY particles \cite{PhysRevD.110.030001}.}
	\label{tab:Sparticles limits}
	\begin{tabular}{lc}
		\hline
		Sparticle &Mass Bounds (GeV) \\
		\hline
		Sleptons& $>107$\\
		Sneutrinos&$>94$\\
		Neutralinos&$>46$\\
		Charginos&$>94$ \\
		\hline
	\end{tabular}
\end{table}
The SUSY particle masses are related to both $m_{1/2}$ and $m_0$. Therefore, we firstly determine their minimum values according to the previous conditions, so $m_{1/2}= 500$ GeV and $m_0=500$ GeV. The parameters space for both tan$\beta$ and $A_0$ are determined by running SPheno without any error. Errors typically occur for two main reasons: (i) Out of the values range of $A_0$ and tan$\beta$, the gauge couplings become large at $M_{GUT}$ due to large beta functions.
\begin{table}[h!tbp]
	\centering
	\caption{Input parameters values in this study.}
	\label{tab:par CMSSM values}
	\begin{tabular}{lc}
		\hline
		Parameter &Values \\
		\hline
		$m_0$ (GeV)& [500,1500]\\
		$m_{1/2}$ (GeV)& [500,1000]\\
		$A_0$ (GeV)&[-1700,1850]\\
		tan$\beta$& [5,20]\\
		cos($\theta_{ij}$)&  [0.087,0.91]\\
		\hline
	\end{tabular}
\end{table}
Hence, the perturbation theory will fail. (ii) Obtaining negative mass square for the SUSY particles \cite{PhysRevD.83.013003}. The values of input parameters are then shown in table \ref{tab:par CMSSM values}.

From equations (\ref{eq:neutrino mass}, \ref{eq:3}, \ref{eq:4}, \ref{eq:5}), we can estimate both values of the Yukawa matrix elements $Y_W$ and the triplet fermion mass $M_W$ according to the mass bound conditions for the light neutrino which is estimated to be $<$ 0.8 eV at low energy scale \cite{Aker2022}. In this study, we consider sizable values for the Yukawa matrix elements. Thus, we should check that they are still within the perturbation regime. Hence, the constraint on the maximum allowed values of entries for the Yukawa matrix is chosen to be: $\mathrm{|}(Y_W)_{ij}\mathrm{|}^2 <4\pi$ \cite{Seesaw4}. Thus, we get $M_W \ge 2.5\times{10}^{13}$ GeV and cos($\theta_{ij}$) $\le$ 0.91. Moreover, sign($\mu)>0$ is fixed for all numerical calculations. From the equations (\ref{eq:3}, \ref{eq:4}, \ref{eq:5}) we have three scenarios for the Yukawa coupling matrix for the $Z$ boson LFV decays. The values of $\left|n_e\right|, \mathrm{|}n_\mu\mathrm{|}$ and $\left|n_\tau\right|$ are set as shown in table \ref{tab:yukawa exampel}.  

\begin{table}[h]
	\centering
	\caption{Yukawa matrix scenarios for numerical calculations of BR\texorpdfstring{($Z\rightarrow l_i l_j$)}{}.}
	\label{tab:yukawa exampel}
	\begin{tabular}{l|llllll|c}
	\hline
	Scenario & $C_{\tau \mu}$ & $C_{\tau e}$  & $C_{\mu e}$ &$\left|n_e\right|$ &$\left|n_\mu \right|$ & $\left|n_\tau \right|$ &\ Yukawa matrix \\
	\hline
	TM &$C_{\tau \mu}$&\ 0 &\ 0 &\ 0.001 & \ 1 &\ 1 & $Y_W = f\left(\begin{matrix}0.001&0&0\\ 0&1&C_{\tau \mu}\\0&0&\sqrt{1-C_{\tau\mu}^2}\\\end{matrix}\right) $ \\
	TE & 0 &$C_{\tau e}$ &\ 0 & \ 1 &\ 0.001 &\ 1&$Y_W = f\left(\begin{matrix}1&0&C_{\tau e}\\0&0.001&0\\ 0&0&\sqrt{1-C_{\tau e}^2}\\\end{matrix}\right) $  \\
	ME & 0 & 0 & $C_{\mu e}$&1&1&0.001& $Y_W = f\left(\begin{matrix}\sqrt{1-C_{\mu e}^2}&0&0\\ C_{\mu e}&1&0\\0&0&0.001\\\end{matrix}\right) $ \\
	\hline	
	\end{tabular}
\end{table}
The energy scale of the GUT is fixed to $M_{GUT} = 2.00\times{10}^{16}$\ GeV. While the SUSY breaking scale is fixed to $M_{SUSY} = {10}^3$\ GeV \cite{Seesaw2}. The BR($Z\rightarrow l_i l_j$) will be studied with(without) applying the constraints due to non-observation of the $l_{i}\rightarrow l_{j} \gamma$ decays.

\subsection{BR($Z\rightarrow l_{i} l_{j}$) without constraints from $(l_{i}\rightarrow l_{j} \gamma)$ decays}
\label{noconstraint}

\begin{figure}[h!tbp]	
	\centering
	\includegraphics[width=.45\textwidth]{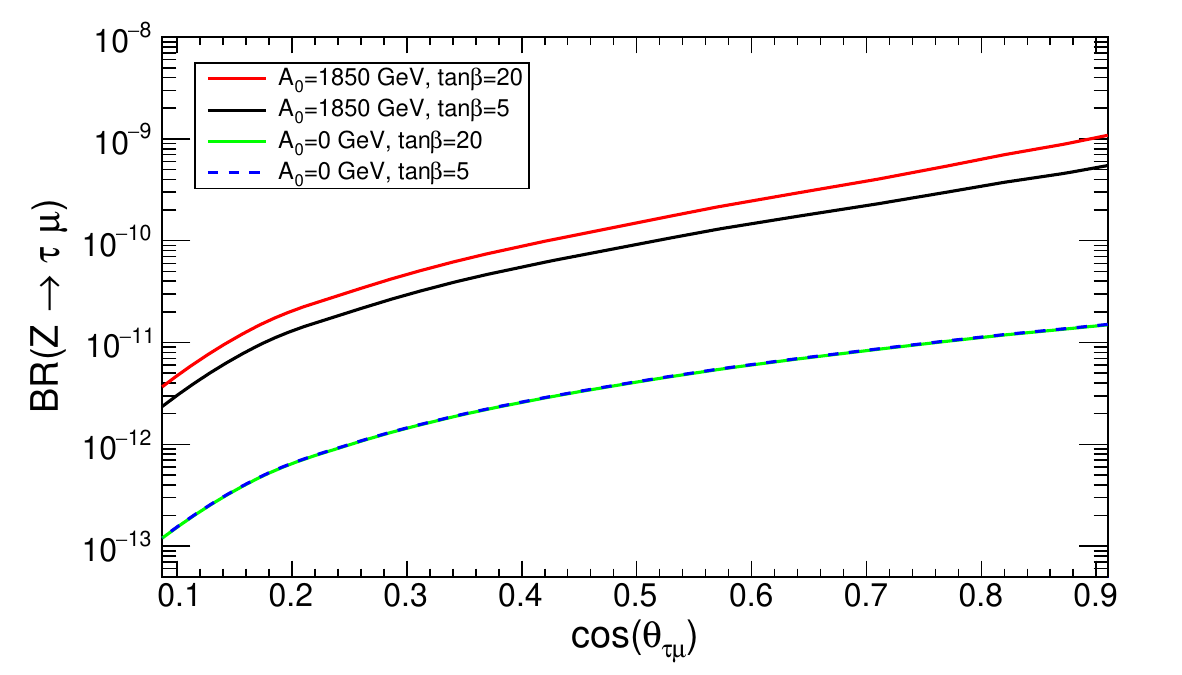}
	\qquad
	\includegraphics[width=.45\textwidth]{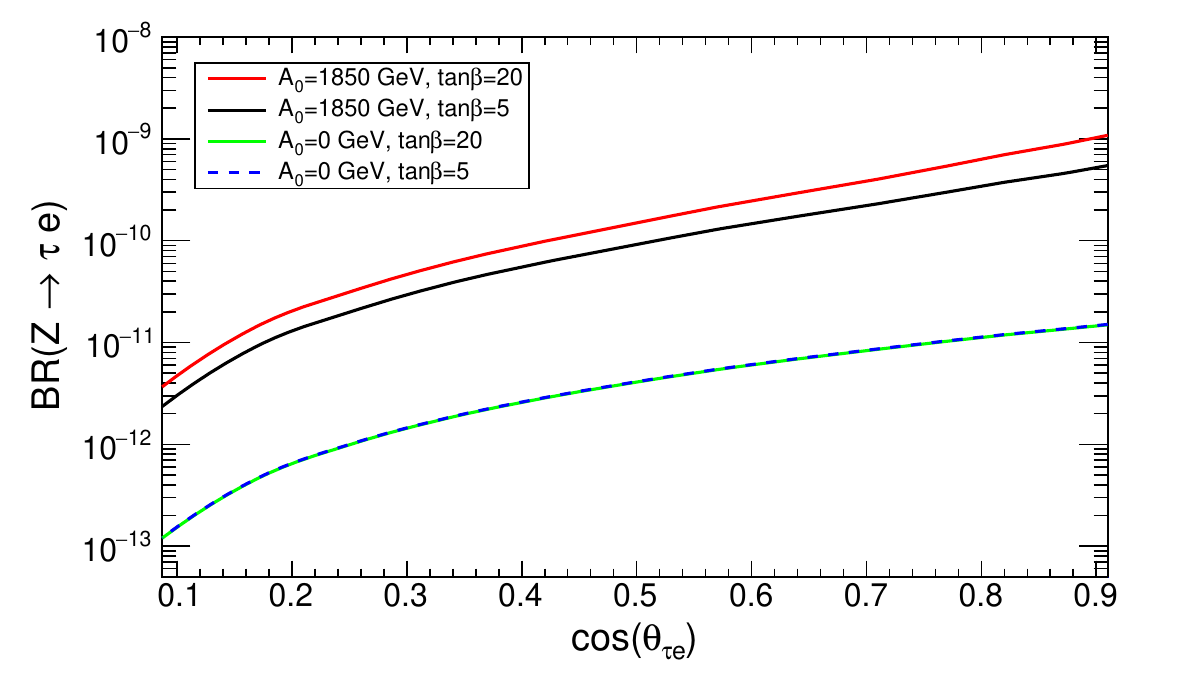}
	\qquad
	\includegraphics[width=.45\textwidth]{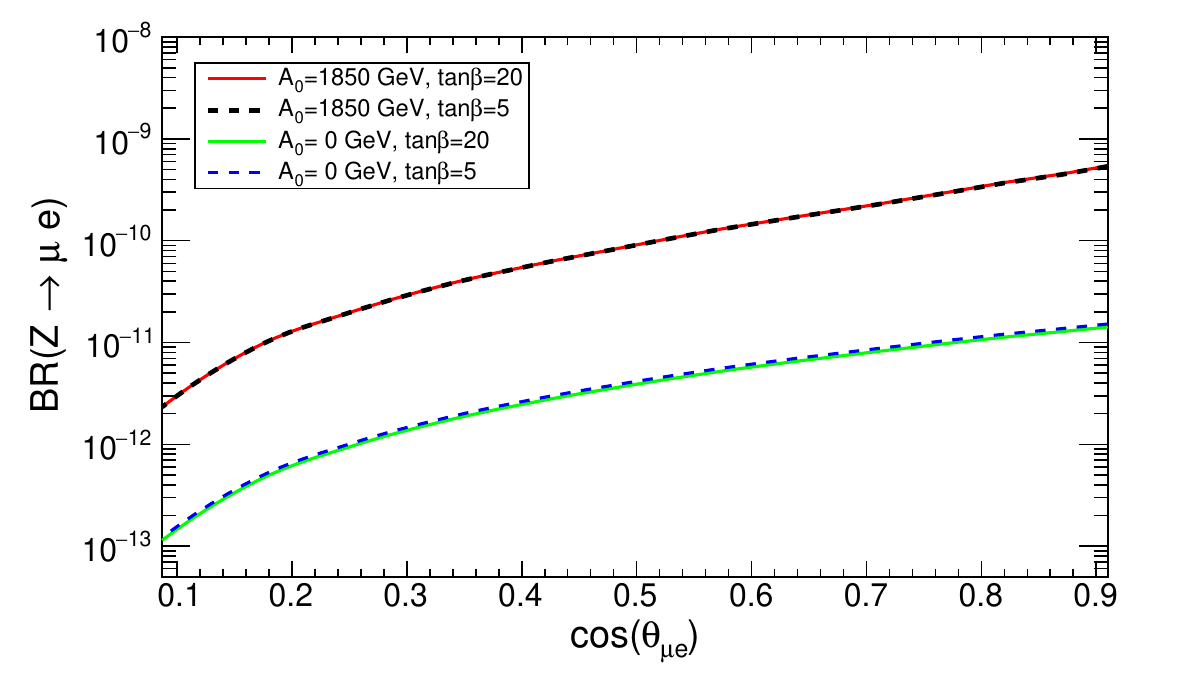}
	\caption{BR$(Z\rightarrow l_i l_j)$ as a function of cos($\theta_{ij}$) at $A_0=0,\ 1850$ GeV and tan$\beta$=5, 20. For all above plots we set $m_{1/2} =$$m_0 = 500$ GeV, $f=1$ and $M_W= 2.5\times{10}^{13}$ GeV. 
		\label{fig:A_0_tanb}}
\end{figure}

In this section, the $Z$ boson LFV decays are considered to be uncorrelated from other LFV decays. We study the BR($Z\rightarrow l_i l_j$) as a function of $A_0$, tan$\beta$, $M_W$, $f$ and $cos(\theta_{ij})$ without constraints from $(l_{i}\rightarrow l_{j} \gamma)$ decays. Fig.\ref{fig:A_0_tanb} shows the BR($Z\rightarrow l_i l_j$) versus cos($\theta_{ij}$) at $A_0=0,\ 1850$ GeV and tan$\beta$=5, 20. We notice from Fig.\ref{fig:A_0_tanb} that the BRs of the $Z$ boson LFV decays increase as cos($\theta_{ij}$) varies from 0.087 to 0.91. For $A_0$ =1850 GeV and tan$\beta$=20, the best values of BR($Z\rightarrow \tau l$) are observed. Whereas, the best value of BR($Z\rightarrow \mu e$) as $A_0$ =1850 GeV and tan$\beta$=5, 20. Furthermore, we notice that BR($Z\rightarrow \mu e$)= BR($Z\rightarrow \tau l$) at two cases, when $A_0$ =0 GeV, tan$\beta$=5, 20 and $A_0=1850$ GeV, tan$\beta$=5. It is obvious that the values of BR($Z\rightarrow l_i l_j$) depend strongly on $A_0$, while their dependence on tan$\beta$ is weak.
 
\begin{figure}[h!tbp]
	\centering
	\includegraphics[width=.45\textwidth]{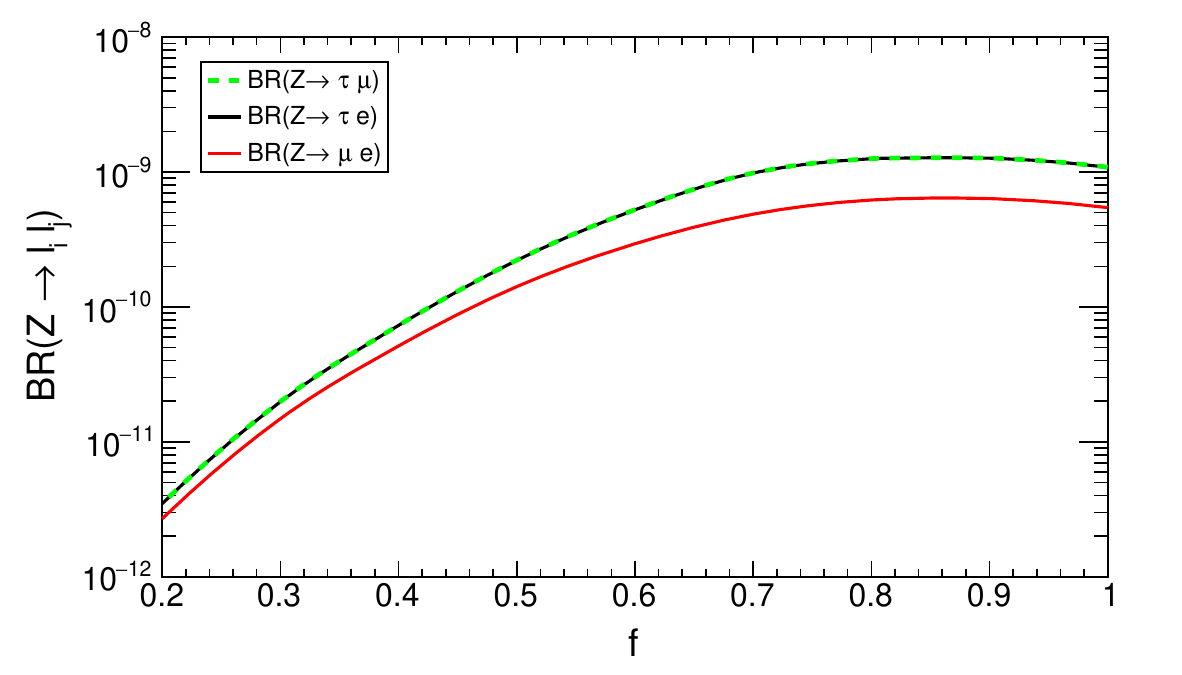}
	\includegraphics[width=.45\textwidth]{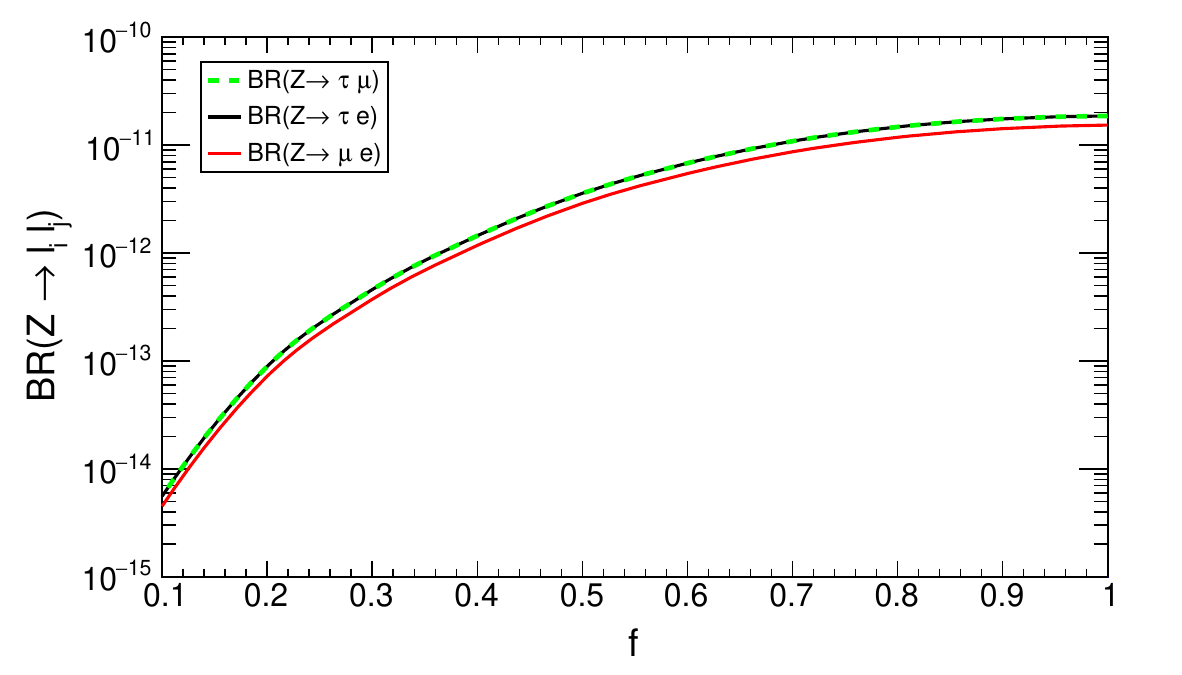}
	\caption{BR$(Z\rightarrow l_i l_j)$ as a function of the strength of neutrino Yukawa coupling (f) at two values of $M_W = 2.5\times{10}^{13}$ GeV (left), and $2.5\times{10}^{14}$ GeV (right). The other input parameter are fixed as $m_0=m_{1/2}=500$ GeV, $A_0=1850$ GeV, tan$\beta$=20 and cos($\theta_{ij}$)=0.91. 
		\label{fig:BR_Z_f+M_W}}
\end{figure}

Additionally, we plot BR$(Z\rightarrow l_i l_j)$ versus strength of the neutrino Yukawa coupling ($f$) for these two values of $M_W = 2.5\times{10}^{13}/2.5\times{10}^{14}$ GeV as shown in Fig.\ref{fig:BR_Z_f+M_W}.

We notice that at $M_W= 2.5\times{10}^{13}$ GeV, the values of BRs of the $Z$ boson LFV decays increase as the $f$ parameter varies from 0.2 to 1. The BRs of $(Z\rightarrow l_i l_j)$ decays reach their maximum values at $f$=0.85, then drop slowly as $f$ becomes larger. The BR($Z\rightarrow\tau\ l$) and BR($Z\rightarrow\mu e$) can be about $1.30\times{10}^{-9}$ and $6.40\times{10}^{-10}$ respectively. Whereas, at $M_W= 2.5\times{10}^{14}$ GeV, the values of the BRs of the $Z$ boson LFV decays increase as the $f$ parameter varies from 0.1 to 1 and their values remain lower than those observed at $M_W= 2.5\times{10}^{13}$ GeV.

\begin{figure}[h!tbp]
	\centering
	\includegraphics[width=.45\textwidth]{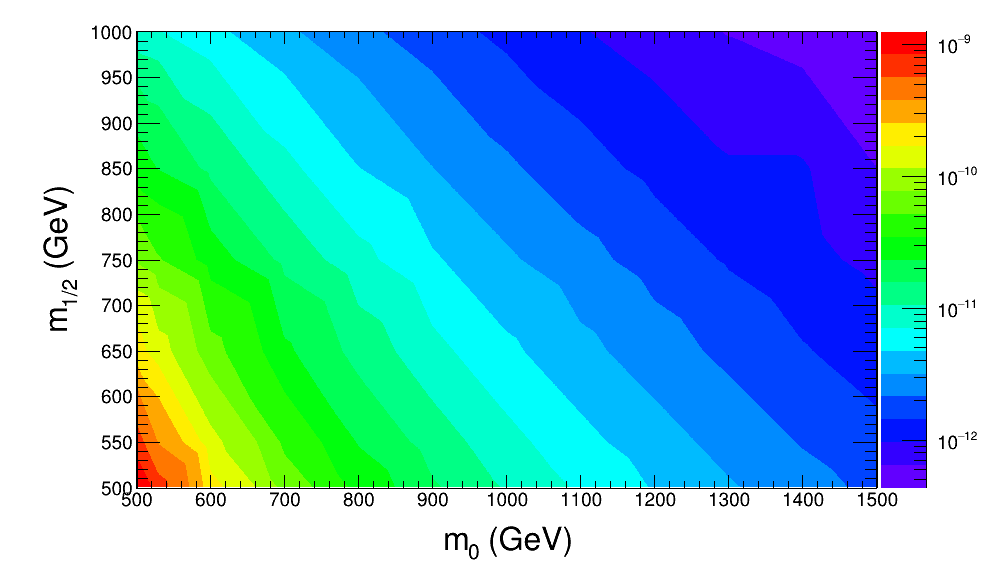}
	\qquad
	\includegraphics[width=.45\textwidth]{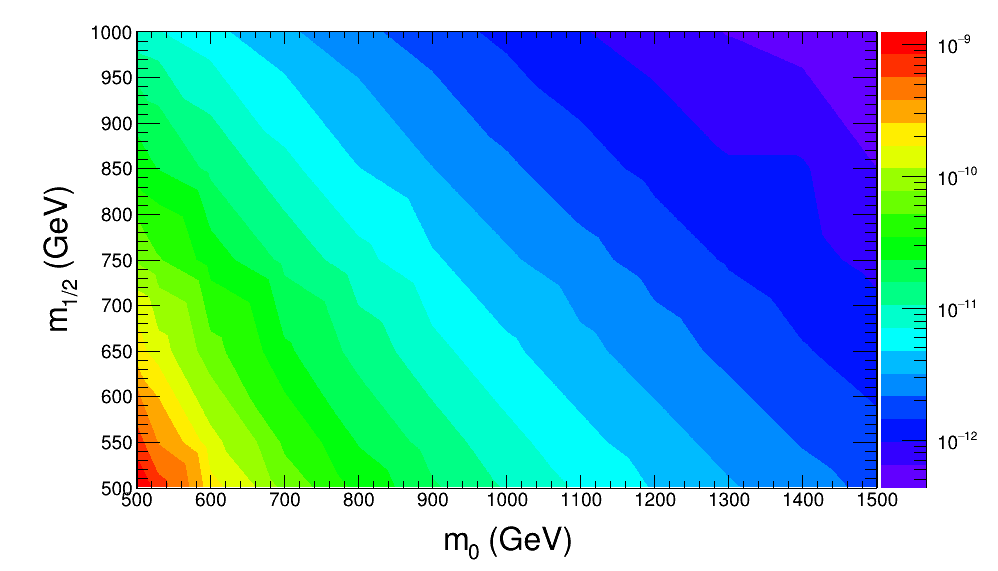}
	\qquad
	\includegraphics[width=.45\textwidth]{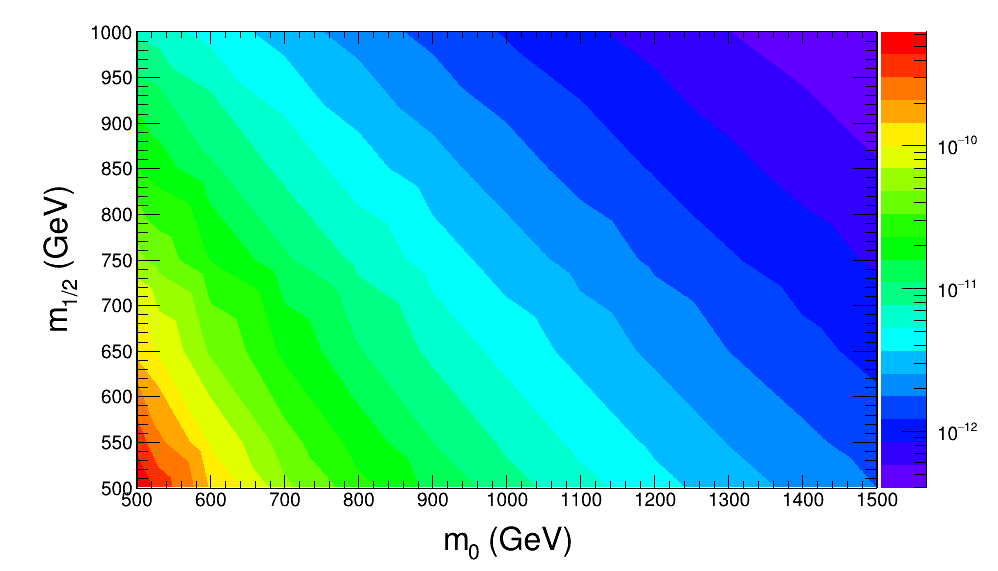}
	\caption{BR$(Z\rightarrow l_i l_j)$ as a contour in the $m_0$ and $m_{1/2}$ plane. BR$(Z\rightarrow \tau \mu)$ shown in upper-left figure, BR$(Z\rightarrow\tau e)$ shown in upper-right one and BR$(Z\rightarrow \mu e)$ shown in bottom one. Other input parameters are fixed as follows: $A_0 = 1850$ GeV, tan$\beta = 20$, cos($\theta_{ij}$) = 0.91, $f$=0.85 and $M_W= 2.5\times{10}^{13}$ GeV. 
		\label{fig:contoure_m_0_m1/2}}
\end{figure}

We also consider the case of the BR($Z\rightarrow l_i l_j$) as a contour in the $m_0$ and $m_{1/2}$ plane as shown in Fig.\ref{fig:contoure_m_0_m1/2}. We notice that the BRs of the $Z$ boson LFV decays decrease as both $m_0$ and $m_{1/2}$ increase. The best values of BRs are at $m_0 = [500,\ 520]$ GeV and $m_{1/2} = [500,\ 530]$ GeV (red region). In this region, the best values of BR($Z\rightarrow\tau l$) and BR($Z\rightarrow \mu e$) are about $1.30\times{10}^{-9}$ and $6.4\times{10}^{-10}$ respectively. 

\subsection{BR($Z\rightarrow l_i l_j$) with constraints from $(l_{i}\rightarrow l_{j} \gamma)$ decays}
In this section, we study the variations of BR($Z\rightarrow l_i l_j$) after applying constraints from Br$(l_{i}\rightarrow l_{j} \gamma)$. The $l_{i}\rightarrow l_{j} \gamma$ decays are not observed. Hence, the BRs for these decays are constrained as shown in table \ref{tab:experimental (l to l+gamma)}.

\begin{table} [h!tbp]
	\centering
	\caption{Experimental upper bounds of the radiative two body decays $(l_{i}\rightarrow l_{j} \gamma)$.}
	\label{tab:experimental (l to l+gamma)}
	\begin{tabular}{lcccc}
		\hline
		Decay & Upper Bound & Experiment\\
		\hline
		$BR(\tau \rightarrow \mu \gamma )$& $4.2\times{10}^{-8}$ \cite{Belle2021} &Belle\\
		
		$BR(\tau \rightarrow e \gamma )$& $5.6\times{10}^{-8}$ \cite{Belle2021} &Belle\\
	
		$BR(\mu \rightarrow e \gamma )$& $4.2\times{10}^{-13}$ \cite{Meg2016} &MEG\\
		\hline
	\end{tabular}
	
\end{table} 
In the supersymmetry, the source of LFV processes can be non-diagonal elements in the left slepton mass matrix. Therefore, in the mass insertion method (MI) with leading-logarithm approximation, the BRs of the $l_{i}\rightarrow l_{j} \gamma$ decays can be approximated as follows \cite{PhysRevD.83.013003, Seesaw3}:
\begin{flalign} 
	\label{eq: (l to l+gamma)}
	BR(l_{i}\rightarrow l_{j} \gamma) \propto \alpha^3 m_{l_i}^5\frac{\left|\Delta m_{\widetilde{L}_{ij}}^2\right|^2}{{\widetilde{m}}^8}{tan}^2(\beta),
\end{flalign} 
where $\alpha$ is the electroweak coupling constant, $m_{l_i}$ is the mass of lepton $i$ and $\widetilde{m}$ is the average of SUSY masses that are involved in loops. From equations (\ref{eq:RGE equations}, \ref{eq: (l to l+gamma)}), we notice that BR$(l_{i}\rightarrow l_{j} \gamma)$ are mainly related to the SUSY mass scale and elements of the non-diagonal left slepton mass matrix.
\begin{figure}[h!tbp]
	\centering
	\begin{subfigure}{0.45\textwidth}
		\includegraphics[width=\textwidth]{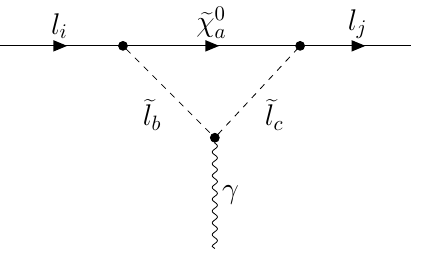}
		\caption{}
	\end{subfigure}
	\hfill
	\begin{subfigure}{0.45\textwidth}
		\includegraphics[width=\textwidth]{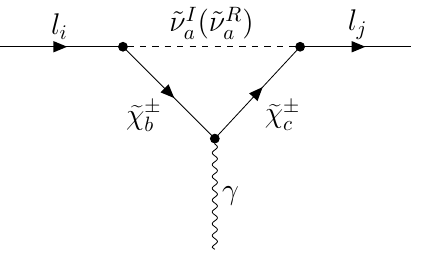}
		\caption{}
	\end{subfigure}
	
	\caption{One loop Feynman diagrams contributing to BR$(l_{i}\rightarrow l_{j} \gamma)$ in the MSSM-Seesaw type-III model. 
		\label{fig: li to lj+gamma}}
\end{figure}
The non-diagonal elements are almost completely governed by the choice of soft SUSY breaking parameters in the heavy seesaw sector. Furthermore, BR$(l_{i}\rightarrow l_{j} \gamma)$ is proportional to the square of tan$\beta$ which leads to larger LFV branching ratios. Therefore, we fixed $A_0=0$ GeV (to exclude tri-linear couplings) and tan$\beta$=5 for all calculations. We also fixed $m_0=m_{1/2}=500$ GeV by considering the conditions of mass bounds for the SUSY particles as shown in table \ref{tab:Sparticles limits}. The LFV of the $l_{i}\rightarrow l_{j} \gamma$ decays is considered to constrain the $f$ parameter, whereas the values of the $Y_W$ matrix elements are fixed as shown in table \ref{tab:yukawa exampel}. The sparticles mediated diagrams for the $l_{i}\rightarrow l_{j} \gamma$ decays in the MSSM-Seesaw type-III model are shown in Fig.\ref{fig: li to lj+gamma}. The off-shell amplitude for $(l_{i}\rightarrow l_{j} \gamma)$ is given by \cite{Porod2014, PhysRevD.53.2442}:
\begin{flalign} 
	M=e\epsilon^{\alpha\ast}{\bar{u}}_i(p-q)\big[q^2\gamma_\alpha(K_1^LP_L+K_1^RP_R)+m_{li}i\sigma_{\alpha\beta}q^\beta(K_2^LP_L+K_2^RP_R) \big]u_j(p),
\end{flalign} 
here $q$ represents the momentum of photon, $e$ is the electric charge, $\epsilon^{\alpha}$ is the polarization vector of photon, $u_i$ and $u_j$ represent the wave function for anti-lepton/lepton and $p$ is momentum of the lepton $li$. In the limit $q \rightarrow 0$, the analytic expression of the BRs of the $(l_{i}\rightarrow l_{j} \gamma)$ decays is:
\begin{flalign} 
	BR(l_{i}\rightarrow l_{j} \gamma)=\alpha \frac{m_{li}^5}{\Gamma_{l_{i}}} (|K_2^L|^2+|K_2^R|^2), 
\end{flalign} 
where $\alpha$ is the fine structure constant and $\Gamma_{l_{i}}$ is the total decay width of $l_{i}$.
$K_2^{L/R}$ are the combinations of the coefficients which correspond to Feynman diagrams as in Fig.\ref{fig: li to lj+gamma} and can be written as:
\begin{flalign} 
	K_2^{L/R}=K_{2a}^{L/R}+K_{2b}^{L/R}.
\end{flalign} 
Here the contributions from neutralino-slepton loops are shown in Fig.\textcolor{blue}{\ref{fig: li to lj+gamma}(a)}. While the contributions from chargino-sneutrino loops are shown in Fig.\textcolor{blue}{\ref{fig: li to lj+gamma}(b)}. The contributions of Fig.\textcolor{blue}{\ref{fig: li to lj+gamma}(a)} are given by:
\begin{flalign} 
	K_2^L \ = \ & 2 V_{\gamma} \big[V_1^L V_2^R  (C_2+C_{12}+C_{11}) m_{e_{{i}}} + V_1^R V_2^L (C_1+C_{12}+C_{11}) m_{e_{{j}}} \nonumber &\\
	& - V_1^R V_2^R  (C_0+C_1+C_2) m_{\tilde{\chi}^0_{{a}}}\big], \\
	K_2^R \ = \ &K_2^L (L\leftrightarrow R). 
\end{flalign} 
The couplings corresponding to Fig.\textcolor{blue}{\ref{fig: li to lj+gamma}(a)} are: $V_{\gamma}=\Gamma^{\tilde{l} \tilde{l}^*\gamma }_{c, b}$, $V_1^{(L/R)}=\Gamma^{\bar{l}\tilde{\chi}^0 \tilde{l} ,{(L/R)}}_{i, a, b}$, and $V_2^{(L/R)}=\Gamma^{\tilde{\chi}^0 l \tilde{l}^*,{(L/R)}}_{a, j, c}$. Table \ref{gamma_vertex} shows the definition of the coupling constants for the Feynman diagrams. 
The contributions of Fig.\textcolor{blue}{\ref{fig: li to lj+gamma}(b)} are given by:
\begin{flalign} 
	K_2^L \ =  \ & -2 (V_\gamma^L)^* V_1^L V_2^R C_{12} m_{e_{{i}}} +2 (V_\gamma^R)^* V_1^R V_2^L (C_2+C_{12}+C_{22}) m_{e_{{j}}} \nonumber &\\
	&+ 2 (V_\gamma^L)^* V_1^R V_2^R C_1 m_{\tilde{\chi}^-_{{b}}} - 2 (V_\gamma^R)^* V_1^R V_2^R (C_0+C_1+C_2) m_{\tilde{\chi}^-_{{c}}}, \\
	K_2^R \ = \ &K_2^L (L\leftrightarrow R).
\end{flalign}
\begin{table} [h!tbp]
	\centering
	\caption{Definition of the coupling constants for the Feynman diagrams. The concrete forms of these couplings are available in the Appendix \ref{section:Appendix-A Vertexes}.}
	\label{gamma_vertex}
	\begin{tabular}{ll}
		\hline
		Coupling Constant& Coupling Definition \\
		\hline \\
		$\Gamma^{\tilde{l} \tilde{l}^*\gamma }_{c, b}$& Coupling of slepton-slepton-gamma vertex \\ \\
		$\Gamma^{\bar{l}\tilde{\chi}^0 \tilde{l} ,{(L/R)}}_{i, a, b}$& Coupling of anti lepton-neutralino-slepton vertex \\ \\
		$\Gamma^{\tilde{\chi}^0 l \tilde{l}^*,{(L/R)}}_{a, j, c}$& Coupling of neutralino-lepton-slepton vertex \\ \\
		$\Gamma^{\tilde{\chi}^+\tilde{\chi}^- \gamma ,{(L/R)}} _{b, c}$& Coupling of chargino-chargino-gamma vertex \\ \\
		$\Gamma^{\bar{l}\tilde{\chi}^- \tilde{\nu}^I ,{(L/R)}}_{i, b, a}$ & Coupling of anti lepton-chargino- CP-odd sneutrino vertex\\ \\
		$\Gamma^{\tilde{\chi}^+ l \tilde{\nu}^I ,{(L/R)}}_{c, j, a}$& Coupling of chargino-lepton- CP-even sneutrino vertex\\ 
		\hline
	\end{tabular}
\end{table}

The couplings corresponding to Fig.\textcolor{blue}{\ref{fig: li to lj+gamma}(b)} are given by: $V_\gamma^{(L/R)}=\Gamma^{\tilde{\chi}^+\tilde{\chi}^- \gamma ,{(L/R)}} _{b, c}$, $V_1^{(L/R)}=\Gamma^{\bar{l}\tilde{\chi}^- \tilde{\nu}^I ,{(L/R)}}_{i, b, a}$, and $V_2^{(L/R)}=\Gamma^{\tilde{\chi}^+ l \tilde{\nu}^I ,{(L/R)}}_{c, j, a}$. The parameters $C_0, C_{00}, C_1, C_2,$ $C_{11}, C_{12}$ and $C_{12}$ denote the standard three-point functions. For Fig.\textcolor{blue}{\ref{fig: li to lj+gamma}(a)} and Fig.\textcolor{blue}{\ref{fig: li to lj+gamma}(b)}, the arguments of the C functions are $(m^2_{\tilde{\chi}^0_{{a}}}, m^2_{\tilde{l}_{{c}}}, m^2_{\tilde{l}_{{b}}})$ and $(m^2_{\tilde{\chi}^-_{{c}}}, m^2_{\tilde{\chi}^-_{{b}}}, m^2_{\tilde{\nu}^I_{{a}}})$ respectively, with vanishing external momenta.
For the CP-even sneutrino couplings, we replace $\tilde{\nu}^I$ with $\tilde{\nu}^R$ in the last two vertices $V_1^{(L/R)}, V_2^{(L/R)}$ and the arguments of C functions.\\

\begin{figure}[h!tbp]
	\centering
	\includegraphics[width=.45\textwidth]{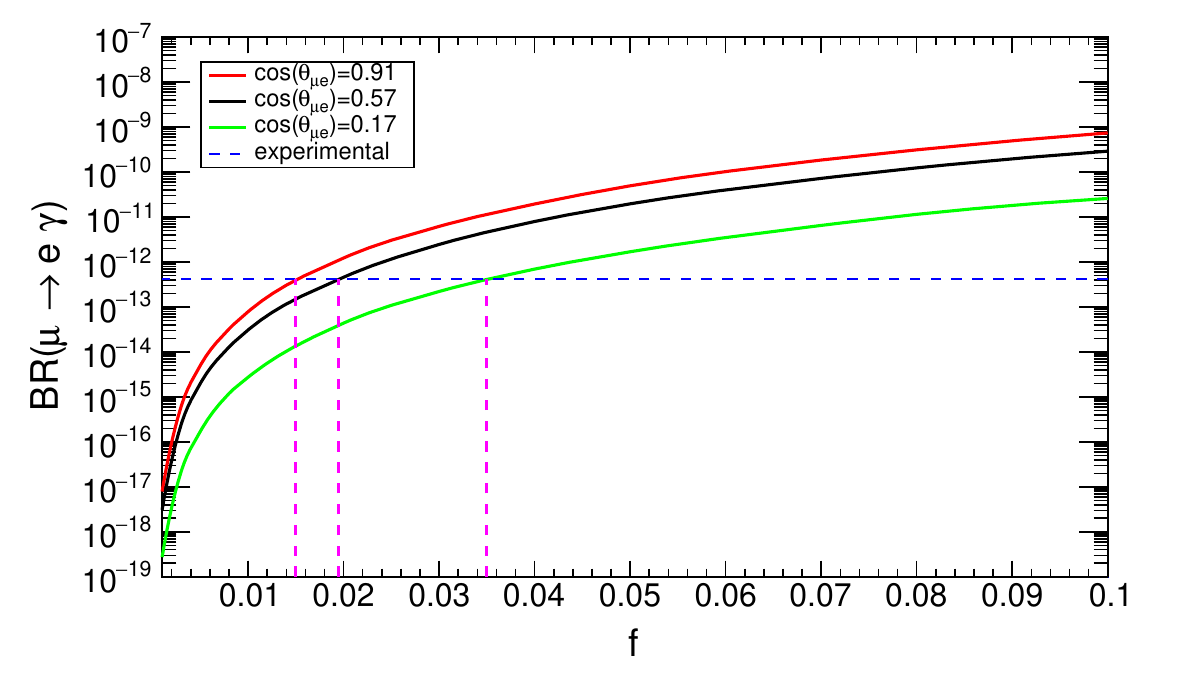}
	\qquad
	\includegraphics[width=.45\textwidth]{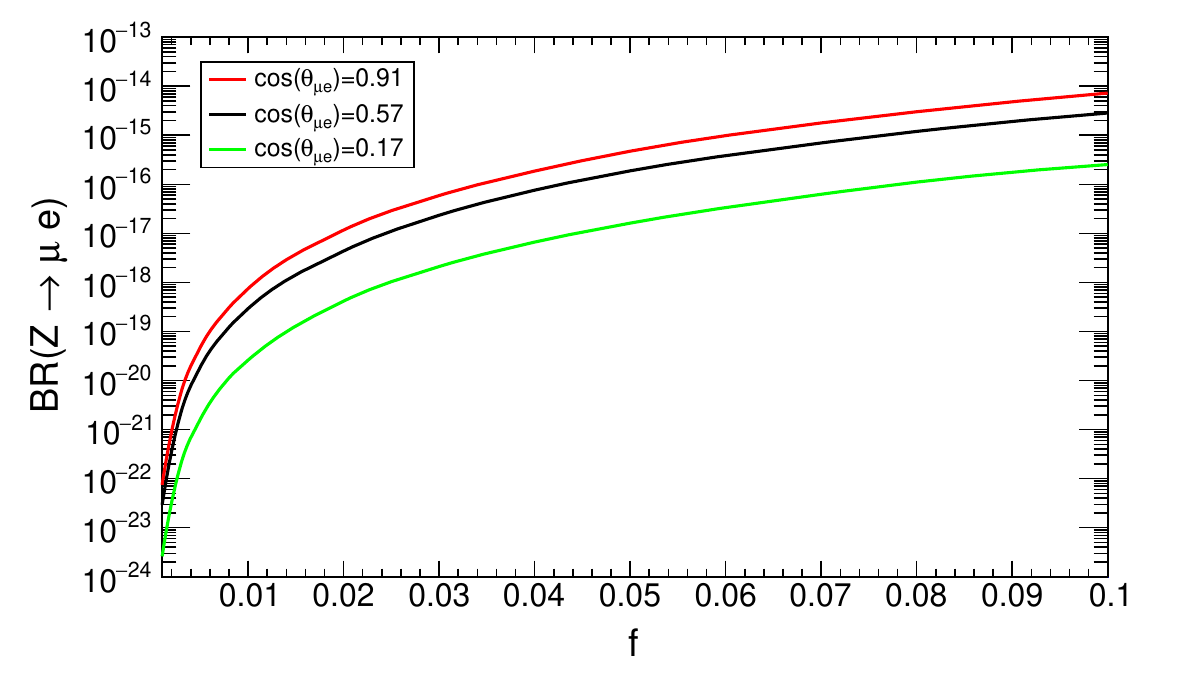}
	\qquad
	\caption{BR$(\mu \rightarrow e \gamma)$ versus $f$ (left), BR$(Z \rightarrow \mu e)$ versus $f$ (right). For these two plots we set cos$(\theta_{\mu e})$=0.91 in red line, cos$(\theta_{\mu e})$=0.57 in black one and cos$(\theta_{\mu e})$=0.17 in green one.
		\label{fig: BR mu_e}}
\end{figure}

\begin{figure}[h!tbp]
	\centering
	\includegraphics[width=.45\textwidth]{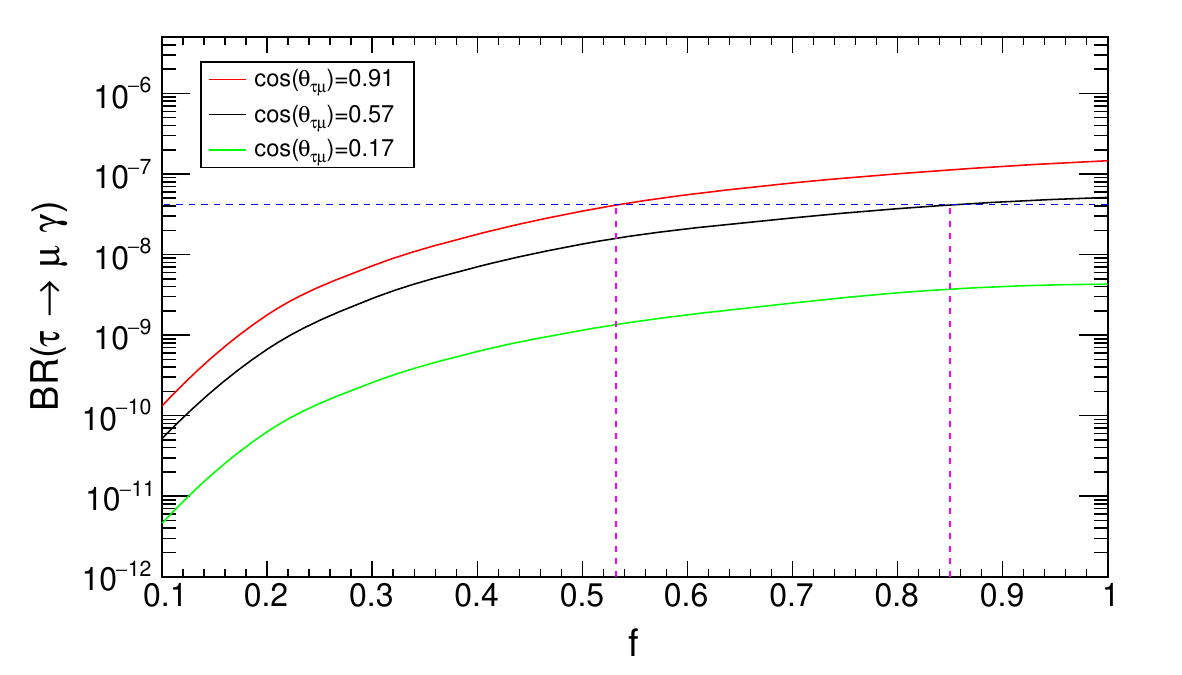}
	\qquad
	\includegraphics[width=.45\textwidth]{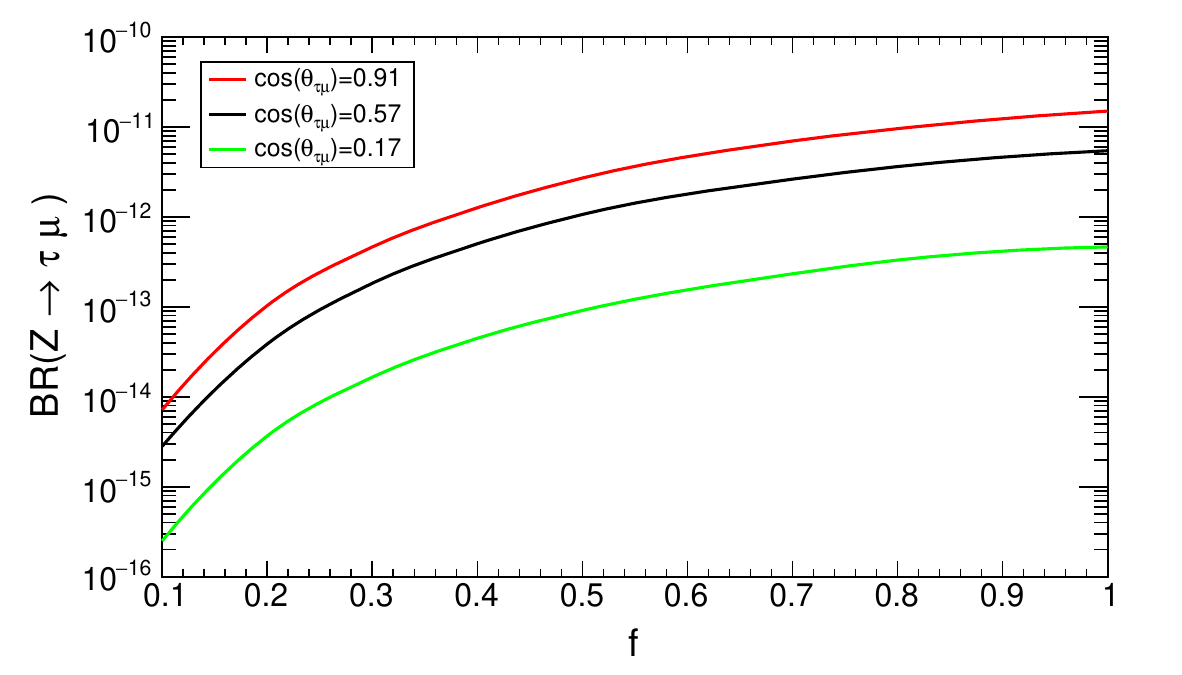}
	\qquad
	\caption{BR$(\tau \rightarrow \mu \gamma)$ versus $f$ (left), BR$(Z \rightarrow \tau \mu)$ versus $f$ (right). For these two plots we set cos$(\theta_{\tau \mu})$=0.91 in red line, cos$(\theta_{\tau \mu})$=0.57 in black one and cos$(\theta_{\tau \mu})$=0.17 in green one. 
		\label{fig: BR mu_tau}}
\end{figure}

\begin{figure}[h!tbp]
	\centering
	\includegraphics[width=.45\textwidth]{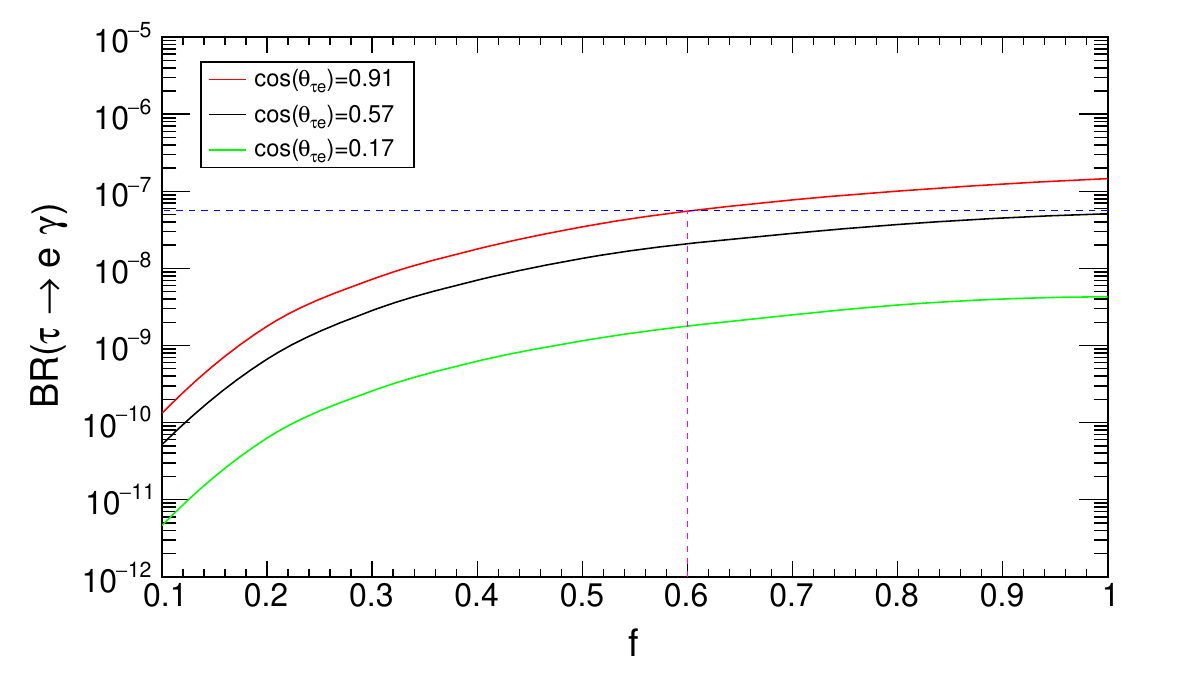}
	\qquad
	\includegraphics[width=.45\textwidth]{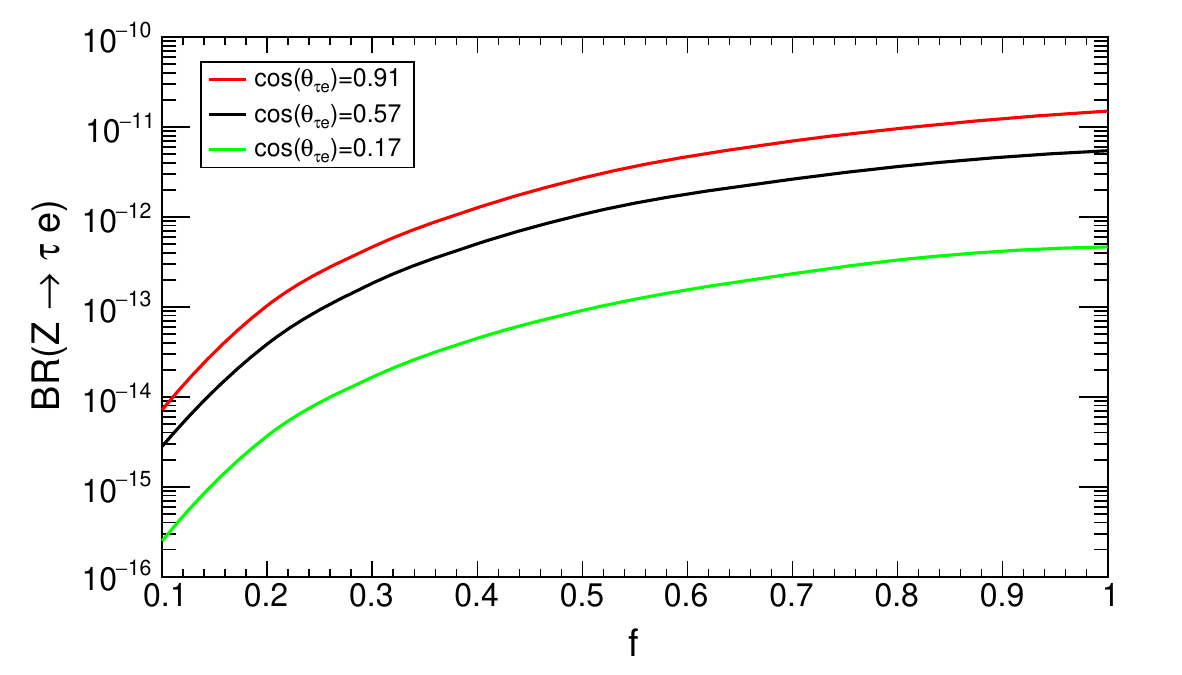}
	\qquad
	\caption{BR$(\tau \rightarrow e\gamma)$ versus $f$ (left), BR$(Z \rightarrow \tau e)$ versus $f$ (right). For these two plots we set cos$(\theta_{\tau e})$=0.91 in red line, cos$(\theta_{\tau e})$=0.57 in black one and cos$(\theta_{\tau e})$=0.17 in green one.  
		\label{fig: BR e_tau}}
\end{figure}

 We calculate the BRs of the $Z$ boson LFV decays by taking $m_0=m_{1/2}=500$ GeV, tan$\beta$=5, $M_W=2.5\times{10}^{13}$ GeV and $A_0=0$ GeV. We plot both BR$(Z \rightarrow l_{i} l_{j})$ and BR$(l_i \rightarrow l_j \gamma)$ versus $f$ at cos$(\theta_{ij})$ =0.91, 0.57 and 0.17 as shown in Fig.(\ref{fig: BR mu_e}, \ref{fig: BR mu_tau}, \ref{fig: BR e_tau}). The horizontal dotted blue line represents the current experimental bound of BR$(l_{i}\rightarrow l_{j} \gamma)$ as shown in table \ref{tab:experimental (l to l+gamma)}. 
We start the numerical discussion with $\mu$ decays since its experimental bounds are the most stringent. From Fig.\ref{fig: BR mu_e}, it is obvious that both BR$(\mu \rightarrow e \gamma)$ and BR$(Z \rightarrow \mu e)$ increase as the $f$ parameter varies from 0.001 to 0.1. The prediction of BR$(\mu \rightarrow e \gamma)$ exceeds the current experimental bound at $f=$0.0151, 0.0195 and 0.035 in red line, black and green one respectively.

So in this case the BR$(Z \rightarrow \mu e)$ is $4.00\times{10}^{-18}$, $4.14\times{10}^{-18}$ and $4.16\times{10}^{-18}$ at $f=$0.015 (cos$(\theta_{\mu e})$=0.91), 0.0195 (cos$(\theta_{\mu e})$=0.57) and 0.035 (cos$(\theta_{\mu e})$=0.17) respectively. Thus, under constraints of the $\mu \rightarrow e \gamma$ decay the value of BR$(Z \rightarrow \mu e)$ is approximately $4.00\times{10}^{-18}$. 

We notice from Fig.\ref{fig: BR mu_tau} that both BR$(\tau \rightarrow \mu \gamma)$ and BR$(Z \rightarrow \tau \mu)$ increase as the $f$ parameter varies from 0.1 to 1. 
The prediction of BR$(\tau \rightarrow \mu \gamma)$ exceeds the current experimental bound at $f=$0.532, 0.85 and 1 in red line, black and green one respectively.
\begin{figure}[h!tbp]
	\centering
	\includegraphics[width=.45\textwidth]{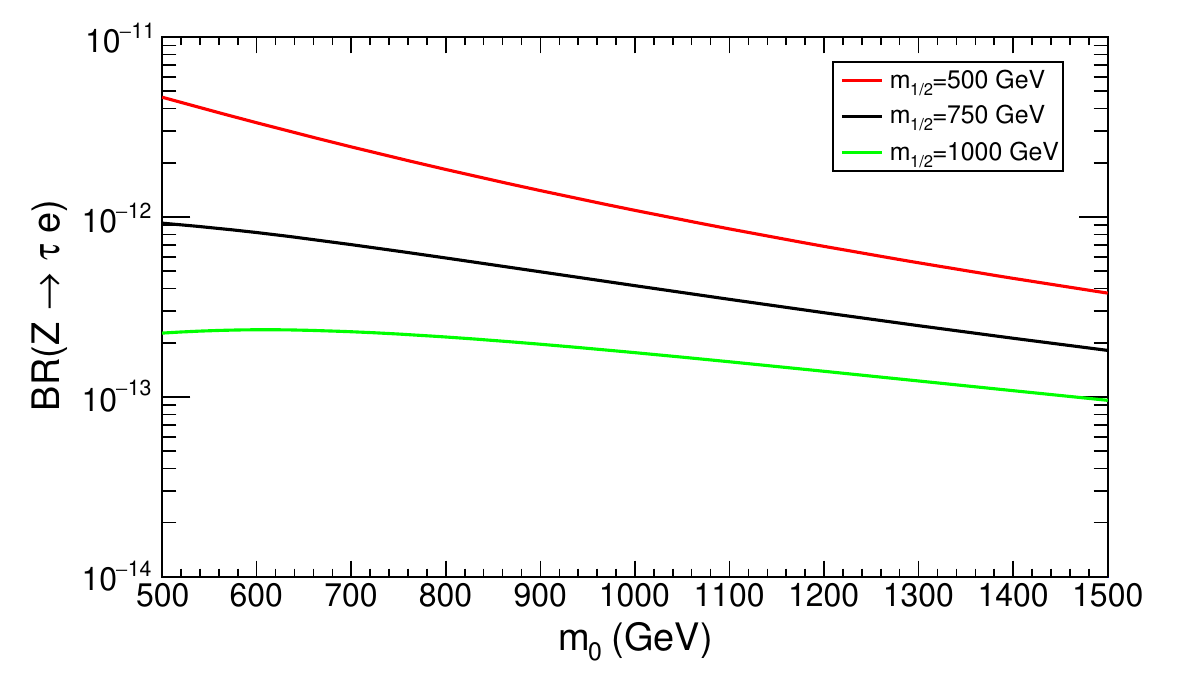}
	\qquad
	\includegraphics[width=.45\textwidth]{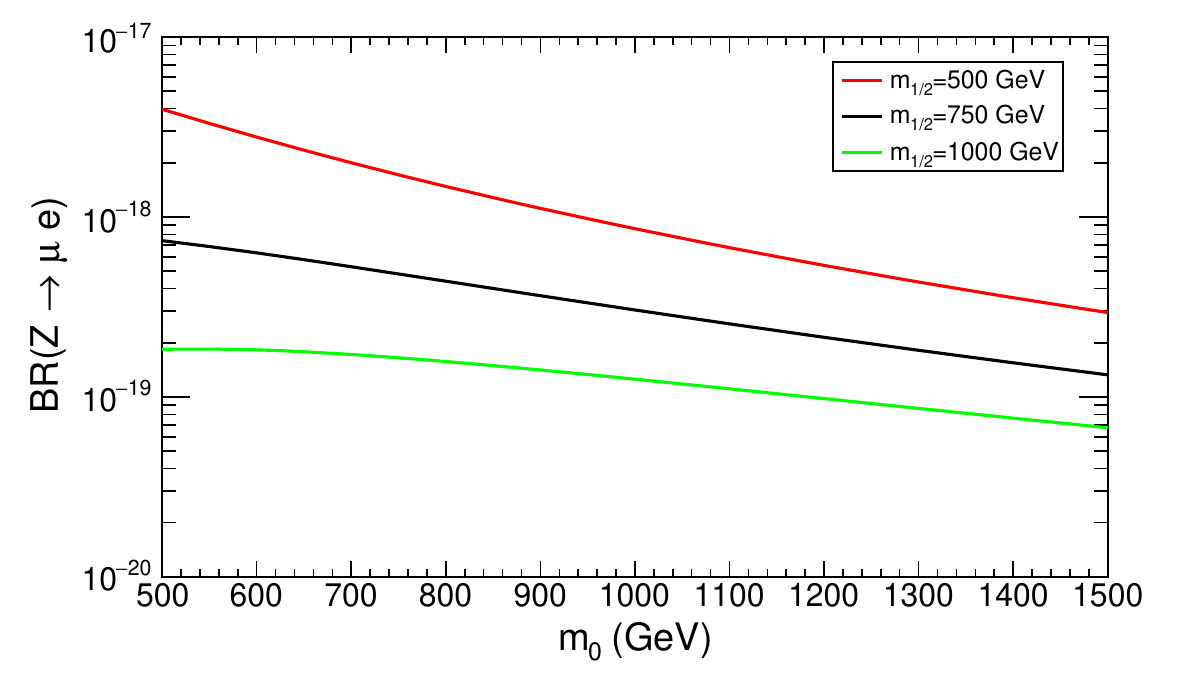}
	\qquad
	\includegraphics[width=.45\textwidth]{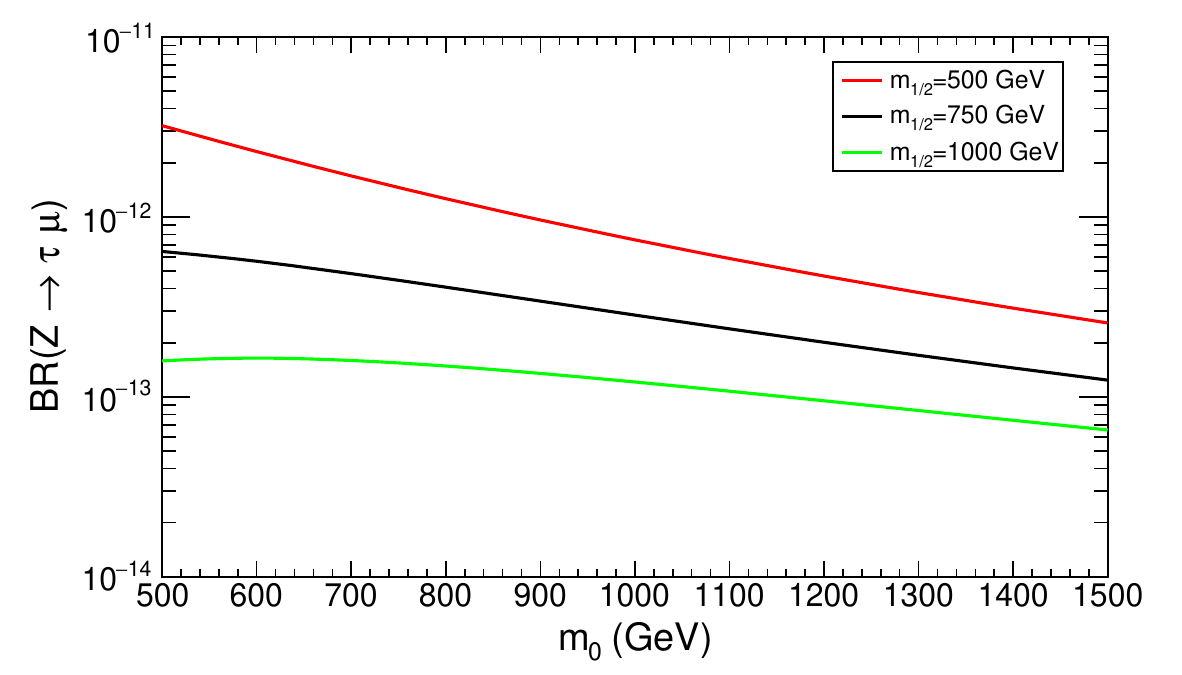}
	\qquad
	\caption{BR$(Z\rightarrow l_i l_j)$ as a function of $m_0$ at $m_{1/2}$=500, 750 and 1000 GeV. For these two plots we set cos$\theta_{ij}$=0.91 at $f$=0.532 for BR$(Z \rightarrow \tau e)$, $f$=0.60 for BR$(Z \rightarrow \tau e)$ and $f$=0.015 for BR$(Z \rightarrow \mu e)$. 
		\label{fig: BRZ constrain}}
\end{figure}

So, in this case the BR$(Z \rightarrow \tau \mu)$ is $3.26\times{10}^{-12}$, $4.00\times{10}^{-12}$ and $4.65\times{10}^{-13}$ at $f=$0.532 (cos$(\theta_{\tau \mu})$=0.91), 0.85 (cos$(\theta_{\tau \mu})$=0.57) and 1 (cos$(\theta_{\tau \mu})$=0.17) respectively. Thus, under constraints of the $\tau \rightarrow \mu \gamma$ decay, the best predicted value of BR$(Z \rightarrow \tau \mu)$ is approximately $4.00\times{10}^{-12}$.

Furthermore, it is clear from Fig.\ref{fig: BR e_tau} that both BR$(\tau \rightarrow e \gamma)$ and BR$(Z \rightarrow \tau e)$ increase as the $f$ parameter varies from 0.1 to 1. The prediction of BR$(\tau \rightarrow e \gamma)$ exceeds the current experimental bound at $f=$0.6, 1, 1 in red line, black and green one respectively. So, in this case the BR$(Z \rightarrow \tau e)$ is $4.61\times{10}^{-12}$, $5.5\times{10}^{-12}$ and $4.65\times{10}^{-13}$ at $f=$0.60 (cos$(\theta_{\tau e})$=0.91), 1 (cos$(\theta_{\tau e})$=0.57) and 1 (cos$(\theta_{\tau e})$=0.17) respectively. Thus, under the constraints of $\tau \rightarrow e \gamma$ decay, the best predicted value of BR$(Z \rightarrow \tau e)$ is approximately $5\times{10}^{-12}$.

Additionally, Fig.\ref{fig: BRZ constrain} shows the variation of BR$(Z\rightarrow l_i l_j)$ as a function of $m_0$ for $m_{1/2}$=500, 750 and 1000 GeV. When $m_0$ and $m_{1/2}$ increase, the BRs of the $Z$ boson LFV decays decrease. Furthermore, under constraints from $l_{i}\rightarrow l_{j} \gamma$ decays, the best values of BR$(Z\rightarrow l_i l_j)$ are obtained at $m_0$=$m_{1/2}=500$ GeV. In this case, BR$(Z \rightarrow \mu e)$ $=4\times{10}^{-18}$, BR$(Z \rightarrow \tau e)$ $=4.6\times{10}^{-12}$ and BR$(Z \rightarrow \tau \mu)$ $=3.2\times{10}^{-12}$. The predicted values are six orders of magnitude below the current experimental bound and four orders of magnitude below the sensitivity of the future experiments.
 
		\section{Conclusions}
The lepton flavor violation (LFV) of the $Z$ boson decays ($Z\rightarrow l_i l_j$) in the constrained MSSM-Seesaw type-III model is studied in this article. In the supersymmetric models, the type-III seesaw mechanism can be realized by adding a fermionic triplet superfield. After deriving the analytical expressions for the branching ratios (BRs) of both $Z$ boson and the radiative two body ($l_{i}\rightarrow l_{j} \gamma$) decays, we have studied numerically how BRs of the $Z$ boson LFV decays can be predicted in this model. The numerical study has been performed by applying the following constraints: fitting to small neutrino masses, conserving the R-parity and the lightest SUSY particle of the considered model is the neutralino. Furthermore, the sparticle masses (charginos, sleptons, sneutrinos and neutralinos) should be above the recent experimental mass bounds. We also applied constraints from the Br$(l_{i}\rightarrow l_{j} \gamma)$ and imposed the perturbativity limits on the parameters of this model.To investigate the small masses of the left-handed neutrinos, the mass of a fermionic triplet $M_W$ should be in the order of $2.5\times{10}^{13}$\ GeV. After satisfying all constraints on the model parameters with(without) constraints from the $l_{i}\rightarrow l_{j} \gamma$ decays, the BRs maximum values of the $Z$ boson LFV decays are summarized as in table \ref{tab: Results}.
\begin{table}[h!tbp]
	\centering
	\caption{Upper bounds of $Z$ boson LFV decays . Our Prediction-1 (Prediction-2) denotes without(with) constraints from the $l_{i}\rightarrow l_{j} \gamma$.}
	\label{tab: Results}
	\begin{tabular}{lcccc}
		\hline
		LFV Decays & Prediction-1 & Prediction-2 &LHC(90\%)& FCC-ee/CEPC\\
		\hline
		BR$(Z\rightarrow \tau \mu)$ & $1.30\times{10}^{-9}$  &$ 3.20\times{10}^{-12}$ & $7.20\times{10}^{-6}$ & ${10}^{-9}$\\
		BR$(Z\rightarrow \tau e)$   &$1.30\times{10}^{-9}$   &$ 4.60\times{10}^{-12}$ & $7.00\times{10}^{-6}$ & ${10}^{-9}$\\
		BR$(Z\rightarrow \mu e)$    &$6.40\times{10}^{-10}$  &$ 4.00\times{10}^{-18}$ & $2.62\times{10}^{-7}$ & $10^{-8}-{10}^{-10}$\\
		\hline
	\end{tabular}
\end{table}

These values are out of the experimental upper bounds for the LHC, while they are in consistent with the sensitivity of the future colliders (FCC-ee/CEPC) for the scenario without constraints from the radiative two body decays $(l_{i}\rightarrow l_{j} \gamma)$ as shown in table \ref{tab: Results}. However, after applying constraints from the $l_{i}\rightarrow l_{j} \gamma$ decays, the above-mentioned results get an additional suppression of about $10^{-3}$ for BR$(Z\rightarrow\tau l)$ and $10^{-8}$ for BR$(Z\rightarrow\mu e)$ when they are compered to the sensitivity of the future colliders (FCC-ee/CEPC). Hence, the BRs predictions are several orders below the recent experimental bounds. 


\bmhead{Acknowledgements}
The authors would like to thank the administration of scientific research at Erzincan Binali Yildirim University/Türkiye, Al-Furat and Idlib University/Syria for supporting and funding this work.

	\appendix
	\appendix

\section{Appendix: Vertexes} \label{section:Appendix-A Vertexes} 
\subsection{Two Fermions - Z Boson Interactions:} 

\begin{align} 
	\Gamma^{\tilde{\chi}^0 \tilde{\chi}^0 Z ,L}_{i, j}=-\frac{i}{2} \Big(g_1 \sin\Theta_W   + g_2 \cos\Theta_W  \Big)\Big(N^*_{j 3} N_{{i 3}}  - N^*_{j 4} N_{{i 4}} \Big)&\\ 
	\Gamma^{\tilde{\chi}^0 \tilde{\chi}^0 Z ,R}_{i, j}= + \,\frac{i}{2} \Big(g_1 \sin\Theta_W   + g_2 \cos\Theta_W  \Big)\Big(N^*_{i 3} N_{{j 3}}  - N^*_{i 4} N_{{j 4}} \Big)\end{align}
\begin{align} 
	\Gamma^{\tilde{\chi}^+\tilde{\chi}^- Z ,L}_{i, j}=\frac{i}{2} \Big(2 g_2 U^*_{j 1} \cos\Theta_W  U_{{i 1}}  + U^*_{j 2} \Big(- g_1 \sin\Theta_W   + g_2 \cos\Theta_W  \Big)U_{{i 2}} \Big)&\\ 
	\Gamma^{\tilde{\chi}^+\tilde{\chi}^- Z ,R}_{i, j}=\frac{i}{2} \Big(2 g_2 V^*_{i 1} \cos\Theta_W  V_{{j 1}}  + V^*_{i 2} \Big(- g_1 \sin\Theta_W   + g_2 \cos\Theta_W  \Big)V_{{j 2}} \Big)\end{align}
\begin{align} 
	\Gamma^{\bar{l} l Z ,L}_{i, j}=\frac{i}{2} \delta_{i j} \Big(- g_1 \sin\Theta_W   + g_2 \cos\Theta_W  \Big)&\\ 
	\Gamma^{\bar{l} l Z ,R}_{i, j}=-i g_1 \delta_{i j} \sin\Theta_W 
\end{align} 

\subsection{Two Scalars - Z Boson Interactions:} 

\begin{align} 
	\Gamma^{\tilde{\nu}^I \tilde{\nu}^R Z }_{i, j}=\frac{1}{2} \Big(g_1 \sin\Theta_W   + g_2 \cos\Theta_W  \Big)\sum_{a=1}^{3}Z^{I,*}_{i a} Z^{R,*}_{j a}
\end{align} 
\begin{align} 
	\Gamma^{\tilde{l} \tilde{l}^*Z } _{i, j}=&\frac{i}{2} \Big(-2 g_1 \sin\Theta_W  \sum_{a=1}^{3}Z^{E,*}_{i 3 + a} Z_{{j 3 + a}}^{E}   + \nonumber &\\ 
	& \Big(- g_1 \sin\Theta_W   + g_2 \cos\Theta_W  \Big)\sum_{a=1}^{3}Z^{E,*}_{i a} Z_{{j a}}^{E}  \Big)
\end{align} 

\subsection{Two Fermions - One Scalar Interactions:} 

\begin{align} 
	\Gamma^{\tilde{\chi}^0 l \tilde{l}^*,L}_{i, j, k}=&i \Big(\frac{1}{\sqrt{2}} g_1 N^*_{i 1} \sum_{a=1}^{3}U^{l,*}_{L,{j a}} Z_{{k a}}^{E}   + \frac{1}{\sqrt{2}} g_2 N^*_{i 2} \sum_{a=1}^{3}U^{l,*}_{L,{j a}} Z_{{k a}}^{E}  \nonumber &\\ 
	&- N^*_{i 3} \sum_{b=1}^{3}U^{l,*}_{L,{j b}} \sum_{a=1}^{3}Y_{l,{a b}} Z_{{k 3 + a}}^{E}   \Big)&\\ 
	\Gamma^{\tilde{\chi}^0 l \tilde{l}^*,R}_{i, j, k}=&+ \,i \Big(- \sqrt{2} g_1 \sum_{a=1}^{3}Z_{{k 3 + a}}^{E} U_{R,{j a}}^{l}  N_{{i 1}}  - \sum_{b=1}^{3}\sum_{a=1}^{3}Y^*_{l,{a b}} U_{R,{j a}}^{l}  Z_{{k b}}^{E}  N_{{i 3}} \Big)
\end{align} 
\begin{align} 
	\Gamma^{\bar{l}\tilde{\chi}^0 \tilde{e} ,L}_{j, i, k}=&i \Big(- N^*_{j 3} \sum_{b=1}^{3}Z^{E,*}_{k b} \sum_{a=1}^{3}U^{l,*}_{R,{i a}} Y_{l,{a b}}    - \sqrt{2} g_1 N^*_{j 1} \sum_{a=1}^{3}Z^{E,*}_{k 3 + a} U^{l,*}_{R,{i a}}  \Big)&\\ 
	\Gamma^{\bar{l}\tilde{\chi}^0 \tilde{l} ,R}_{j, i, k}=&+ \,i \Big(\frac{1}{\sqrt{2}} \sum_{a=1}^{3}Z^{E,*}_{k a} U_{L,{i a}}^{l}  \Big(g_1 N_{{j 1}}  + g_2 N_{{j 2}} \Big) - \sum_{b=1}^{3}\sum_{a=1}^{3}Y^*_{l,{a b}} Z^{E,*}_{k 3 + a}  U_{L,{i b}}^{l}  N_{{j 3}} \Big)
\end{align} 

\begin{align} 
	\Gamma^{\bar{l}\tilde{\chi}^- \tilde{\nu}^I ,L}_{i, j, k}=&- \frac{1}{\sqrt{2}} U^*_{j 2} \sum_{b=1}^{3}Z^{I,*}_{k b} \sum_{a=1}^{3}U^{l,*}_{R,{i a}} Y_{l,{a b}}\\
	\Gamma^{\bar{l}\tilde{\chi}^- \tilde{\nu}^I ,R}_{i, j, k}=& + \,\frac{1}{4} \Big(2 \sqrt{2} g_2 \sum_{a=1}^{3}Z^{I,*}_{k a} U_{L,{i a}}^{l}  V_{{j 1}}  \nonumber &\\ 
	&+ v_u \Big(\sum_{b=1}^{3}Z^{I,*}_{k b} \sum_{a=1}^{3}\kappa^*_{{\nu},{a b}} U_{L,{i a}}^{l}   + \sum_{b=1}^{3}\sum_{a=1}^{3}\kappa^*_{{\nu},{a b}} Z^{I,*}_{k a}  U_{L,{i b}}^{l} \Big)V_{{j 2}} \Big)
\end{align} 
\begin{align} 
	\Gamma^{\bar{l}\tilde{\chi}^- \tilde{\nu}^R ,L}_{i, j, k}=&i \frac{1}{\sqrt{2}} U^*_{j 2} \sum_{b=1}^{3}Z^{R,*}_{k b} \sum_{a=1}^{3}U^{l,*}_{R,{i a}} Y_{l,{a b}} \\
	\Gamma^{\bar{l}\tilde{\chi}^- \tilde{\nu}^R ,R}_{i, j, k}= &+ \,\frac{i}{4} \Big(-2 \sqrt{2} g_2 \sum_{a=1}^{3}Z^{R,*}_{k a} U_{L,{i a}}^{l}  V_{{j 1}}   \nonumber& \\ 
	&+ v_u \Big(\sum_{b=1}^{3}Z^{R,*}_{k b} \sum_{a=1}^{3}\kappa^*_{{\nu},{a b}} U_{L,{i a}}^{l}   + \sum_{b=1}^{3}\sum_{a=1}^{3}\kappa^*_{{\nu},{a b}} Z^{R,*}_{k a}  U_{L,{i b}}^{l} \Big)V_{{j 2}} \Big)
\end{align} 
\begin{align} 
	\Gamma^{\tilde{\chi}^+ l \tilde{\nu}^I ,L}_{i, j, k}=&\frac{1}{4} \Big(-2 \sqrt{2} g_2 V^*_{i 1} \sum_{a=1}^{3}U^{l,*}_{L,{j a}} Z^{I,*}_{k a}  \nonumber &\\ 
	& - v_u V^*_{i 2} \Big(\sum_{b=1}^{3}Z^{I,*}_{k b} \sum_{a=1}^{3}U^{l,*}_{L,{j a}} \kappa_{\nu,{a b}}   + \sum_{b=1}^{3}U^{l,*}_{L,{j b}} \sum_{a=1}^{3}Z^{I,*}_{k a} \kappa_{\nu,{a b}}  \Big)\Big)\\
	\Gamma^{\tilde{\chi}^+ l \tilde{\nu}^I ,R}_{i, j, k}=& + \,\frac{1}{\sqrt{2}} \sum_{b=1}^{3}Z^{I,*}_{k b} \sum_{a=1}^{3}Y^*_{l,{a b}} U_{R,{j a}}^{l}   U_{{i 2}}
\end{align} 
\begin{align} 
	\Gamma^{\tilde{\chi}^+ l \tilde{\nu}^R ,L}_{i, j, k}=&\frac{i}{4} \Big(-2 \sqrt{2} g_2 V^*_{i 1} \sum_{a=1}^{3}U^{l,*}_{L,{j a}} Z^{R,*}_{k a}  + v_u V^*_{i 2} \Big(\sum_{b=1}^{3}Z^{R,*}_{k b} \sum_{a=1}^{3}U^{l,*}_{L,{j a}} \kappa_{\nu,{a b}}\nonumber   & \\
	&+ \sum_{b=1}^{3}U^{l,*}_{L,{j b}} \sum_{a=1}^{3}Z^{R,*}_{k a} \kappa_{\nu,{a b}}  \Big)\Big)\\
	\Gamma^{\tilde{\chi}^+ l \tilde{\nu}^R ,R}_{i, j, k}=& + \,i \frac{1}{\sqrt{2}} \sum_{b=1}^{3}Z^{R,*}_{k b} \sum_{a=1}^{3}Y^*_{l,{a b}} U_{R,{j a}}^{l}   U_{{i 2}}
\end{align} 

\subsection{Two Scalars - $\gamma$ Interactions:}

\begin{align} 
	\Gamma^{\tilde{l} \tilde{l}^*\gamma }_{i, j}=\frac{i}{2} \Big(2 g_1 \cos\Theta_W  \sum_{a=1}^{3}Z^{E,*}_{i 3 + a} Z_{{j 3 + a}}^{E}   + \Big(g_1 \cos\Theta_W   + g_2 \sin\Theta_W  \Big)\sum_{a=1}^{3}Z^{E,*}_{i a} Z_{{j a}}^{E}  \Big)
\end{align} 

\subsection{Two Fermions - $\gamma$ Interactions:} 

\begin{align} 
	\Gamma^{\tilde{\chi}^+\tilde{\chi}^- \gamma ,{(L)}} _{i, j}=\frac{i}{2} \Big(2 g_2 U^*_{j 1} \sin\Theta_W  U_{{i 1}}  + U^*_{j 2} \Big(g_1 \cos\Theta_W   + g_2 \sin\Theta_W  \Big)U_{{i 2}} \Big)\\ 
	\Gamma^{\tilde{\chi}^+\tilde{\chi}^- \gamma ,{(R)}} _{i, j}=\frac{i}{2} \Big(2 g_2 V^*_{i 1} \sin\Theta_W  V_{{j 1}}  + V^*_{i 2} \Big(g_1 \cos\Theta_W   + g_2 \sin\Theta_W  \Big)V_{{j 2}} \Big)
\end{align}

\bibliography{sn-bibliography}%

\end{document}